\documentclass[aps,pre,superscriptaddress,nofootinbib,twocolumn,10pt]{revtex4-2}

\usepackage[utf8]{inputenc} 
\usepackage[T1]{fontenc}    
\usepackage{hyperref}       
\usepackage{url}            
\usepackage{booktabs}       
\usepackage{amssymb,amsmath,amsfonts, dsfont}       
\usepackage{nicefrac}       
\usepackage{graphicx}       
\graphicspath{{./grafici/}} 
\usepackage{float} 
\usepackage{algorithm} 
\usepackage{algpseudocode} 
\usepackage{physics}
\usepackage{xcolor}
\usepackage{enumitem}
\usepackage{xcolor}

\newcommand{\bea}{\begin{eqnarray}}
\newcommand{\eea}{\end{eqnarray}}

\newcommand{\<}{\langle}
\renewcommand{\>}{\rangle}

\newcommand{\HH}{\mathcal{H}}
\newcommand{\PGB}{P_\text{GB}}

\newcommand{\Nf}{\text{4N}}
\newcommand{\s}{\sigma}
\newcommand{\bs}{\boldsymbol{\sigma}}

\newcommand{\changes}[1]{\textcolor{black}{#1}}

\begin{document}

\title{Nearest-Neighbours Neural Network architecture\\ for efficient sampling of statistical physics models}

\author{Luca Maria Del Bono}\thanks{Corresponding author: \href{mailto:lucamaria.delbono@uniroma1.it}{lucamaria.delbono@uniroma1.it}}
\affiliation{Dipartimento di Fisica, Sapienza Università di Roma, Piazzale Aldo Moro 5, Rome 00185, Italy}
\affiliation{CNR-Nanotec, Rome unit, Piazzale Aldo Moro 5, Rome 00185, Italy}

\author{Federico Ricci-Tersenghi}
\affiliation{Dipartimento di Fisica, Sapienza Università di Roma, Piazzale Aldo Moro 5, Rome 00185, Italy}
\affiliation{CNR-Nanotec, Rome unit, Piazzale Aldo Moro 5, Rome 00185, Italy}
\affiliation{ INFN, sezione di Roma1, Piazzale Aldo Moro 5, Rome 00185, Italy}

\author{Francesco Zamponi}
\affiliation{Dipartimento di Fisica, Sapienza Università di Roma, Piazzale Aldo Moro 5, Rome 00185, Italy}

\begin{abstract}
The task of sampling efficiently the Gibbs-Boltzmann distribution of disordered systems is important both for the theoretical understanding of these models and for the solution of practical optimization problems.
Unfortunately, this task is known to be hard, especially for spin-glass-like problems at low temperatures.
Recently, many attempts have been made to tackle the problem by mixing classical Monte Carlo schemes with newly devised Neural Networks that learn to propose smart moves. In this article, we introduce the Nearest-Neighbours Neural Network ($\Nf$) architecture, a physically interpretable deep architecture whose number of parameters scales linearly with the size of the system and that can be applied to a large variety of topologies. We show that the $\Nf$ architecture can accurately learn the Gibbs-Boltzmann distribution for a prototypical spin-glass model, the two-dimensional Edwards-Anderson model, and specifically for some of its most difficult instances. In particular, it captures properties such as the energy, the correlation function and the overlap probability distribution. 
Finally, we show that the $\Nf$ performance increases with the number of layers, in a way that clearly connects to the correlation length of the system, thus providing a simple and interpretable criterion to choose the optimal depth. 
\end{abstract}

\maketitle

\section*{Introduction}


A central problem in many fields of science,
from statistical physics to computer science
and artificial intelligence, is that of sampling
from a complex probability distribution over a large
number of variables.
More specifically, a very common such probability
is a Gibbs-Boltzmann distribution. 
Given a set of $N\gg 1$ random variables $\bs = \{\s_1,\cdots,\s_N\}$, 
such a distribution is written in the form
\begin{equation}\label{eq:GB_distribution}
    \PGB(\bs) = \frac{e^{-\beta \HH(\bs)}}{\mathcal{Z}(\beta)} \ .
\end{equation}
In statistical physics language,
the normalization constant $\mathcal{Z}(\beta)$ is called partition function, $\HH(\bs)$ is called
the Hamiltonian function 
and $\beta = 1/T$ is the inverse temperature, corresponding to a global rescaling
of~$\HH(\bs)$.

While Eq.~\eqref{eq:GB_distribution} is simply a trivial definition, i.e. $\HH(\bs) = - \frac1{\beta}\log \PGB(\bs) + \text{const}$,
the central idea of Gibbs-Boltzmann distributions is
that $\HH(\bs)$ is expanded as a sum of `local' interactions. For instance, 
in the special case of binary (Ising) variables, $\s_i \in \{-1,+1\}$, one can always write
\begin{equation}\label{eq:H_exp}
    \HH = - \sum_{i} H_i \sigma_i 
    -\sum_{ i<j} J_{ij} \sigma_i \sigma_j 
    - \sum_{i<j<k} J_{ijk} \sigma_i \sigma_j \sigma_k + \cdots
\end{equation}
In many applications, like in physics (i.e. spin glasses), inference (i.e. maximum entropy models) and artificial intelligence (i.e. Boltzmann machines), the expansion in Eq.~\eqref{eq:H_exp} is truncated to the pairwise terms, thus neglecting higher-order interactions. This leads to a Hamiltonian
\begin{equation}\label{eq:spin_glass_H}
    \HH = - \sum_{i} H_i \sigma_i 
    -\sum_{ i<j} J_{ij} \sigma_i \sigma_j 
\end{equation}
parametrized by `local \changes{external} fields' $H_i$ and `pairwise couplings' $J_{ij}$.
In physics applications such as spin glasses, these
are often chosen to be \changes{independent and identically} distributed random variables, e.g.
$H_i \sim \mathcal{N}(0,H^2)$ and
$J_{ij} \sim \mathcal{N}(0,J^2/N)$.
In Boltzmann Machines, instead, the fields and couplings are learned by maximizing the likelihood of a given training set.

In many cases, dealing with Eq.~\eqref{eq:GB_distribution} analytically,
e.g.\ computing expectation values of interesting quantities, is impossible, and one resorts to numerical computations. A universal strategy is to use {\it local} Markov Chain Monte Carlo (MCMC) methods, which have the advantage of being applicable to a wide range of different systems. 
In these methods, one proposes a local move, typically flipping a single spin, $\sigma_i \to -\sigma_i$, and accepts or rejects the move in such a way to guarantee that after many iterations, the configuration $\bs$ is distributed according to Eq.~\eqref{eq:GB_distribution}.
Unfortunately, these methods are difficult (if not altogether impossible) to apply in many hard-to-sample problems for which the convergence time is very large, which, in practice, ends up requiring a huge amount of computational time.
For instance, finding the ground state of the Sherrington-Kirkpatrick model (a particular case of sampling at $T\to 0$) is known to be a NP-hard
problem; sampling a class of optimization problems is known to take a time scaling exponentially with $N$. The training of Boltzmann Machines is also known to suffer from this kind of problems.

These inconveniences can be avoided by using system-specific {\it global} algorithms that leverage properties of the system under examination in order to gain a significant speedup. Instead of flipping one spin at a time, these methods are able to construct smart moves, which update simultaneously a large number of spins.
An example is the Wolff algorithm \cite{wolff_collective_1989} for the Ising model.
Another class of algorithms uses unphysical moves or extended variable spaces, such as the Swap Monte Carlo for glass-forming models of particles \cite{Ninarello2017}. 
The downside of these algorithms is that they cannot generally be transferred from one system to another, meaning that one has to develop new techniques specifically for each system of interest.
Yet another class is Parallel Tempering (PT) \cite{Marinari}, or one of its modifications \cite{houdayer_cluster_2001}, which considers a group of systems at different temperatures and alternates between simple MCMC moves and moves that swap the systems at two different temperatures. While PT is a universal strategy, its drawback is that, in order to produce low-temperature configurations, one has to simulate the system at a ladder of higher temperatures, which can be computationally expensive and redundant.


The rise of machine learning technology has sparked a new line of research aimed at using Neural Networks to enhance Monte Carlo algorithms, in the wake of similar approaches in many-body quantum physics \cite{carleo_solving_2017}, molecular dynamics \cite{tamagnone2024coarse} and the study of glasses \cite{jung2023roadmap, galliano2024policy, ciarella2023finding}. The key idea is to use the network to propose new states with a probability $P_\text{NN}$ that (i)~can be efficiently sampled and (ii)~is close to the Gibbs-Boltzmann distribution, e.g., with respect to the Kullback-Leibler (KL) divergence $D_{\text{KL}}$:
\begin{equation}\label{eq_DKL}
    D_{\text{KL}}(P_{\text{GB}} \parallel P_{\text{NN}}) = \sum_{\boldsymbol{\sigma}} P_{\text{GB}}(\boldsymbol{\sigma}) \log \frac{P_{\text{GB}}(\boldsymbol{\sigma})}{P_{\text{NN}}(\boldsymbol{\sigma})} \ .
\end{equation}
The proposed configurations can be used in a  Metropolis scheme, accepting them with probability:
\begin{equation}\label{eq:prob}
\begin{split}
    \text{Acc}\left[\boldsymbol{\sigma} \rightarrow \boldsymbol{\sigma}'\right] &= \min\left[1, \frac{P_\text{GB}(\boldsymbol{\sigma}') \times P_{\text{NN}}(\boldsymbol{\sigma})}{P_\text{GB}(\boldsymbol{\sigma}) \times P_{\text{NN}}(\boldsymbol{\sigma}')}\right]\\
    &= \min\left[1, \frac{e^{-\beta \HH (\boldsymbol{\sigma}')} \times P_{\text{NN}}(\boldsymbol{\sigma})}{e^{-\beta \HH (\boldsymbol{\sigma})} \times P_{\text{NN}}(\boldsymbol{\sigma}')}\right].
    \end{split}
\end{equation}
Because $\mathcal{H}$ can be computed in polynomial time and 
$P_{\text{NN}}$ can be sampled efficiently by hypothesis, this approach ensures an efficient sampling of the Gibbs-Boltzmann probability, as long as (i) $P_{\text{NN}}$ covers well the support of $P_{\text{GB}}$ (i.e., the so-called mode collapse is avoided) and (ii) the acceptance probability in Eq.~\eqref{eq:prob} is significantly different from zero for most of the generated configurations. This scheme can also be combined with standard local Monte Carlo moves to improve efficiency~\cite{gabrie_adaptive_2022}.

The main problem is how to train the Neural Network to minimize Eq.~\eqref{eq_DKL}; in fact, the computation of $D_{\rm KL}$ requires sampling from
$P_{\text{GB}}$. A possible solution is to use
$D_{\text{KL}}(P_{\text{NN}} \parallel P_{\text{GB}})$ instead~\cite{wu_solving_2019}, but this has been shown
to be prone to mode collapse~\cite{ciarella_machine-learning-assisted_2023}.
Another proposed solution is to set up an iterative procedure, called `Sequential Tempering' (ST), in which one first learns $P_{\text{NN}}$ at high temperature where sampling from $P_{\text{GB}}$ is possible, then
gradually decreases the temperature, updating $P_{\text{NN}}$ slowly~\cite{mcnaughton_boosting_2020}. 
A variety of new methods and techniques have thus been proposed to tackle the problem, such as autoregressive models~\cite{wu_solving_2019,mcnaughton_boosting_2020,trinquier2021efficient}, 
normalizing flows \cite{kohler_equivariant_2020, gabrie_adaptive_2022, dibak_temperature_2022, invernizzi2022skipping}, and diffusion models \cite{chen2024diffusive}.
It is still unclear whether this kind of strategies actually performs well on hard problems and challenges already-existing algorithms~\cite{ciarella_machine-learning-assisted_2023}. 
The first steps in a theoretical understanding of the performances of such algorithms are just being made \cite{ghio2023sampling,kilgour2022inside}. The main drawback of these architectures is that
the number of parameters scales poorly with
the system size, typically as $N^\alpha$ with 
$\alpha>1$, while local Monte Carlo requires a number 
of operations scaling linearly in $N$.
Furthermore, these architectures -- with the notable
exception of TwoBo~\cite{biazzo_sparse_2024} that directly inspired our work -- are not informed about the physical properties of the model, and in particular about the structure of its interaction graph and correlations.
It should be noted that the related problem of finding the ground state of the system (which coincides with zero-temperature sampling) has been tackled using similar ideas, again with positive~\cite{schuetz_combinatorial_2022, schuetz_graph_2022, fan_searching_2023,pugacheva2024enhancing,colantonio2024efficient} and negative results~\cite{angelini_modern_2022, boettcher2023inability,boettcher2024deep,fan2023reply}.

In this paper, we introduce the Nearest Neighbours Neural Network, or $\Nf$ for short, a Graph Neural Network-inspired architecture that implements an autoregressive scheme to learn the Gibbs-Boltzmann distribution. This architecture has a number of parameters scaling linearly in the system size and can sample new configurations in $\mathcal{O}(N)$ time, thus achieving the best possible scaling one could hope for such architectures. Moreover, $\Nf$ has a straightforward physical interpretation. In particular, the choice of the number of layers can be directly linked to the correlation length of the model under study. Finally, the architecture can easily be applied to essentially any statistical system, such as lattices in higher dimensions or random graphs, and it is thus more general than other architectures.
As a proof of concept, we evaluate the $\Nf$ architecture on the two-dimensional Edwards-Anderson spin glass model, a standard benchmark also used in previous work~\cite{mcnaughton_boosting_2020,ciarella_machine-learning-assisted_2023}. 
We demonstrate that the model succeeds in accurately learning the Gibbs-Boltzmann distribution, especially for some of the most challenging model instances. Notably, it precisely captures properties such as energy, the correlation function, and the overlap probability distribution.

\section*{State of the art}

Some common architectures used for $P_{\text{NN}}$ are normalizing flows~\cite{kohler_equivariant_2020, gabrie_adaptive_2022, dibak_temperature_2022, invernizzi2022skipping} and diffusion models~\cite{chen2024diffusive}.
In this paper, we will however focus on autoregressive models~\cite{wu_solving_2019}, which make use of the factorization:
\begin{equation}\label{eq:factorization}
\begin{split}
&    P_\text{GB}(\boldsymbol{\sigma}) = P(\sigma_1) P(\sigma_2 \mid \sigma_1) P(\sigma_3 \mid \sigma_1, \sigma_2) \times \\ & \times \cdots P(\sigma_n \mid \sigma_{1}, \cdots, \sigma_{n-1}) = \prod_{i=1}^{N} P(\sigma_i \mid \boldsymbol{\sigma}_{<i}) \ ,
\end{split}
\end{equation}
where \(\boldsymbol{\sigma}_{<i} = (\sigma_1, \sigma_2, \ldots, \sigma_{i-1})\),
so that states can then be generated using \textit{ancestral sampling}, i.e. generating first $\sigma_1$, then $\sigma_2$ conditioned on the sampled value of $\sigma_1$, then $\sigma_3$ conditioned to $\sigma_1$ and $\sigma_2$ and so on. The factorization in~\eqref{eq:factorization} is exact, but computing $P(\s_i \mid \bs_{<i})$ exactly generally requires a number of operations scaling exponentially with $N$. Hence, in practice, 
$P_\text{GB}$ is approximated by $P_\text{NN}$ that takes the form in Eq.~\eqref{eq:factorization} where each individual term $P(\s_i \mid \bs_{<i})$ is approximated using a small set of parameters. 
Note that the unsupervised problem of approximating $P_\text{GB}$ is now formally translated into a supervised problem of learning the output, i.e. the probability $\pi_i \equiv P(\sigma_i=+1 \mid \bs_{<i})$, as a function of the input $\bs_{<i}$.
The specific way in which this approximation is carried out depends on the architecture.
In this section, we will describe some common autoregressive architectures found in literature,
for approximating the function~$\pi_i(\bs_{<i})$.

\begin{itemize}
    \item  In the \textbf{Neural Autoregressive Distribution Estimator (NADE)} architecture \cite{mcnaughton_boosting_2020}, the input $\bs_{<i}$ is encoded into a vector $\mathbf{y}_i$ of size $N_h$ using
\begin{equation}
\mathbf{y}_i = \Psi \left( \underline{\underline{A}} \boldsymbol{\sigma}_{<i} + \mathbf{B} \right),
\end{equation}
where $\mathbf{y}_i \in \mathbb{R}^{N_h}$, $\underline{\underline{A}} \in \mathbb{R}^{N_h \times N}$ and $\boldsymbol{\sigma}_{<i}$ is the vector of the spins in which the spins $l \ge i$ have been masked, i.e. $\mathbf{\sigma}_{<i} = (\sigma_1, \sigma_2, \dots, \sigma_{i-1}, 0, \dots, 0 )$. $\mathbf{B} \in \mathbb{R}^{N_h}$ is the bias vector and $\Psi$ is an element-wise activation function. Note that $\underline{\underline{A}}$ and $\mathbf{B}$ do not depend on the index $i$, i.e. they are shared across all spins. The information from the hidden layer is then passed to a fully connected layer
\begin{equation}
\changes{\psi_i} = \Psi \left( \mathbf{V}_i \cdot \mathbf{y}_i + C_i \right) 
\end{equation}
where $\mathbf{V}_i \in \mathbb{R}^{N_h}$ and $C_i \in \mathbb{R}$. Finally, the output probability is obtained by applying a sigmoid function to the output, $\pi_i = S(\changes{\psi_i})$. The Nade Architecture uses $\mathcal{O}(N\times N_h)$ parameters. However, the choice of the hidden dimension $N_h$ and how it should be related to $N$ is not obvious.
For this architecture to work well in practical applications, 
$N_h$ needs to be $\mathcal{O}(N)$,
thus leading to $\mathcal{O}(N^2)$ parameters~\cite{mcnaughton_boosting_2020}.

\item The \textbf{Masked Autoencoder for Distribution Estimator (MADE)} \cite{germain2015made} is a generic dense architecture with the addition of the autoregressive requirement, obtained by setting, for all layers $l$ in the network, the weights $W^{l}_{ij}$ between node $j$ at layer $l$ and node $i$ at layer $l+1$ equal to 0 when $i \ge j$. Between one layer and the next one adds nonlinear activation functions and the width of the hidden layers can also be increased. The MADE architecture, despite having high expressive power, has at least $\mathcal{O}(N^2)$ parameters, which makes it poorly scalable.

\item \textbf{Autoregressive Convolutional Neural Network (CNN)} architectures, such as PixelCNN~\cite{van2016conditional,madanchi2024simulations}, are networks that implement a CNN structure but superimpose the autoregressive property. These architectures typically use translationally invariant kernels ill-suited to study disordered models.

\item The \textbf{TwoBo (two-body interaction)} architecture \cite{biazzo_sparse_2024} is derived from the observation that, given the Gibbs-Boltzmann probability in Eq.~\eqref{eq:GB_distribution} and the Hamiltonian in Eq.~\eqref{eq:spin_glass_H}, the probability $\pi_i$ can be exactly rewritten as \cite{biazzo2023autoregressive}
\begin{equation}\label{eq:twoboP}
\pi_i(\boldsymbol{\sigma}_{<i}) = S \left( 2\beta h^i_{i} + 2\beta H_i + \rho_i(\boldsymbol{h}^i) \right),
\end{equation}
where $S$ is the sigmoid function, $\rho_i$ is a highly non-trivial function and the 
$\boldsymbol{h}^i = \{ h^i_i, h^i_{i+1}, \cdots h^i_N \}$ 
are defined as:
\begin{equation}\label{eq:hilTB}
    h^i_{l} = \sum_{f<i} J_{lf} \sigma_f \quad \text{for} \quad l \geq i.
\end{equation}
Therefore, the TwoBo architecture first computes the vector $\boldsymbol{h}^i$, 
then approximates $\rho_i(\boldsymbol{h}^i)$ using a Neural Network (for instance, a Multilayer Perceptron) and finally computes Eq.~\eqref{eq:twoboP} using a skip connection (for the term $2\beta h^i_{i}$) and a sigmoid activation. By construction, in a two-dimensional model with $N=L^2$ variables and nearest-neighbor interactions, for given $i$ only $\mathcal{O}(L=\sqrt{N})$ of the $h^i_{l}$ are non-zero, see Ref.~\cite[Fig.1]{biazzo_sparse_2024}. 
Therefore, the TwoBo architecture has
$\mathcal{O}(N^\frac{3}{2})$ parameters in the
two-dimensional case.
However, the scaling becomes worse in higher dimensions $d$, i.e. $\mathcal{O}(N^\frac{2d-1}{d})$ and becomes the same as MADE (i.e. $\mathcal{O}(N^2)$) for random graphs.
Additionally, the choice of the $h^i_l$, although justified mathematically, does not have a clear physical interpretation: the TwoBo architecture takes into account far away spins which, however, should not matter much when $N$ is large.

\end{itemize}

To summarize, the NADE, MADE and TwoBo architectures all need $\mathcal{O}(N^2)$ parameters for generic models defined on random graph structures. TwoBo has less parameters for models defined on $d$-dimensional lattices, but still
$\mathcal{O}(N^\frac{2d-1}{d}) \gg \mathcal{O}(N)$.
CNN architectures with translationally invariant kernels have fewer parameters, but are not suitable for disordered systems\changes{, are usually applied to $d=2$ and remain limited to $d \leq 3$}.

To solve these problems, we present here the new,
physically motivated $\Nf$ architecture that has $\mathcal{O}(\mathcal{A} N)$ 
parameters, with a prefactor $\mathcal{A}$ that remains finite
for $N\to\infty$ and is interpreted as the correlation length of the system, as we show in the next section.

\begin{figure*}[t] 
    \centering
    \makebox[\textwidth][c]{\includegraphics[width=0.8\textwidth]{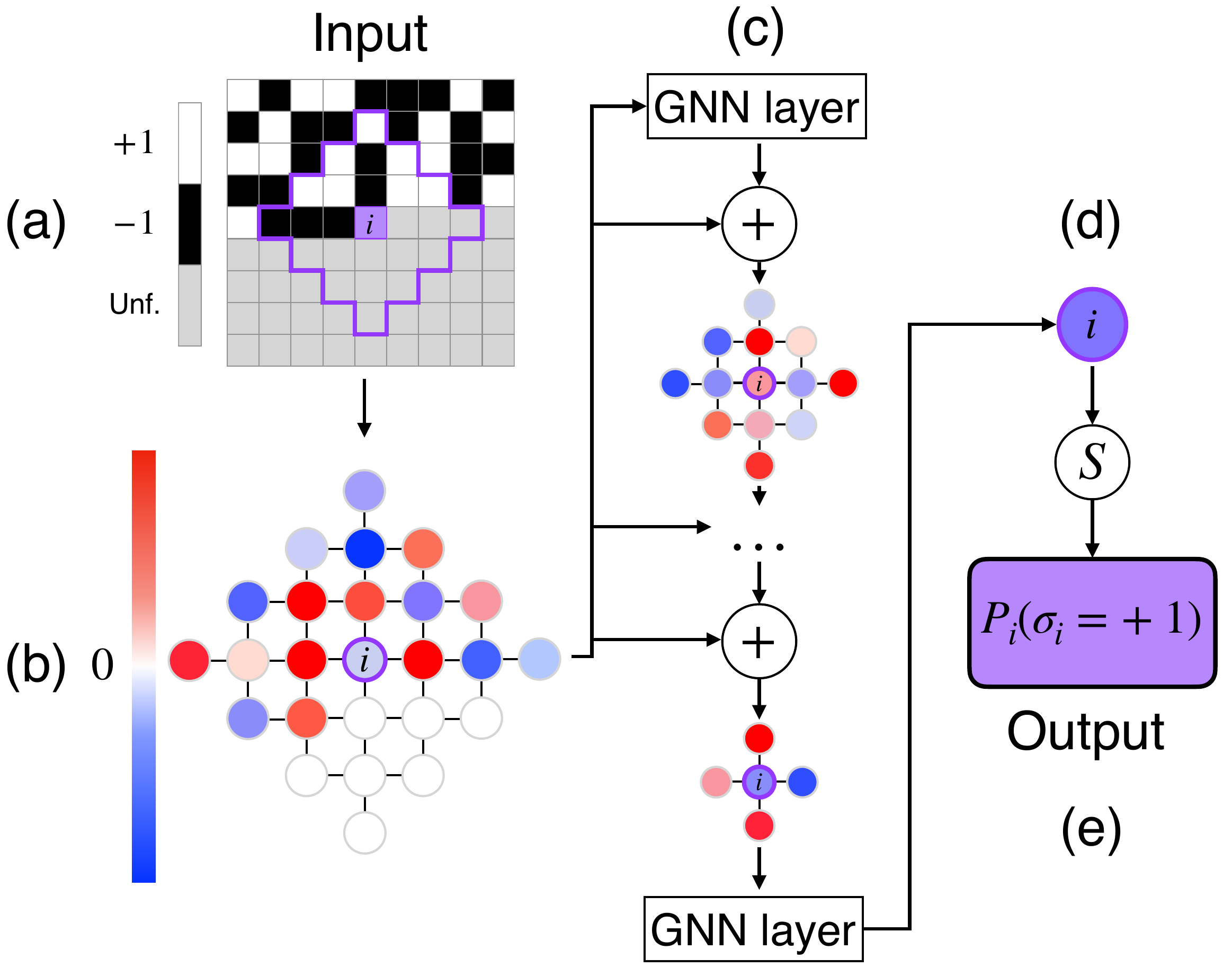}}%
    \caption{Schematic representation of the $\Nf$ architecture implemented on a two-dimensional lattice model with nearest-neighbors interactions. 
    (a)~We want to estimate the probability $\pi_i \equiv P(\sigma_i=+1 \mid \bs_{<i})$ that the spin $i$ (in purple) has value $+1$ given the spins $< i$ (in black and white). The spins $> i$ (in grey) have not yet been fixed.
    (b)~We compute the values of the local fields $h_l^{i,(0)}$ as in Eq.~\eqref{eq:hilTB}, for $l$ in a neighborhood of $i$ of size $\ell$, indicated by a purple contour in (a).  Notice that spins $l$ that are not in the neighborhood of one of the fixed spin have $h_l^{i,(0)} = 0$ and are indicated in white in (b). (c)~In the main cycle of the algorithm, we apply a series of $\ell$ GNN layers with the addition of skip connections.  (d)~The final GNN layer is not followed by a skip connection and yields the final values of the field $h_i^{i,(\ell)}$. (e) $\pi_i$ is estimated by applying a sigmoid function to $h_i^{i,(\ell)}$. Periodic (or other) boundary conditions can be considered, but are not shown here for clarity.}
    \label{fig:N4scheme}
\end{figure*}

\section*{Results}\label{sec:numerics}

\subsection*{The $\Nf$ architecture}

The $\Nf$ architecture (Fig. \ref{fig:N4scheme}) computes the probability 
$\pi_i \equiv P(\sigma_i=+1 \mid \bs_{<i})$ 
by propagating the $h_l^i$, defined as in TwoBo, Eq.~\eqref{eq:hilTB}, through a Graph Neural Network (GNN) architecture.
The variable $h_l^i$ is interpreted as the local field induced by the frozen variables $\bs_{<i}$ on spin $l$. Note that in $\Nf$, we consider all possible values of $l=1,\cdots,N$, not only $l\geq i$ as in TwoBo.
The crucial difference with TwoBo is that only 
the fields in a finite neighborhood of variable $i$ need to be considered, because
the initial values of the fields are propagated through a series of GNN layers that take into account the locality of the physical problem.
During propagation, fields are combined with layer- and edge-dependent weights together with the couplings $J_{ij}$ and the inverse temperature $\beta$.
After each layer of the network (except the final one), a skip connection to the initial configuration is added. After the final layer, a sigmoidal activation is applied to find~$\pi_i$. 
\changes{The network has a total number of parameters equal to $(c+1) \times \ell \times N$, where $c$ is the average number of neighbors for each spin.} Details are given in the Methods section.

More precisely,
because GNN updates are local, and there are $\ell$ of them, in order to compute the final field on \changes{site $i$, hence $\pi_i$, we only need to consider the set of fields corresponding to sites at distance at most $\ell$, hence the number of operations scales proportionally to the number of such sites which we call $\mathcal{B}(\ell)$. Because we need to compute $\pi_i$ for each of the $N$ sites, the generation of a new configuration then requires $\mathcal{O}(\mathcal{B}(\ell) \times N)$ steps.
We recall that in a $d-$dimensional lattice $\mathcal{B}(\ell)\propto \ell^d$, while for random graphs $\mathcal{B}(\ell) \propto c^\ell$.} It is therefore crucial to assess how the number of layers $\ell$ should scale with the system size $N$.

\begin{figure}[t] 
    \centering
\includegraphics[width=
\columnwidth]{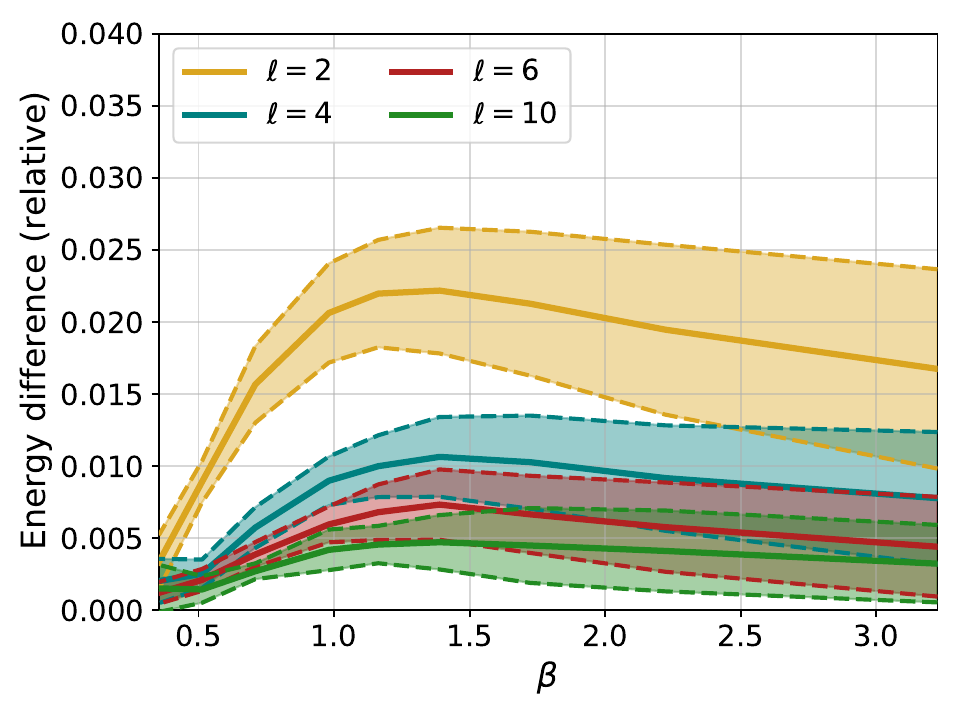}%
    \caption{Relative difference in mean energy between the configurations generated by the $\Nf$ architecture and by PT, for different numbers of layers $\ell$ and different values of the inverse temperature $\beta$. Solid lines indicate averages over 10 different instances, while the dashed lines and the colored area identify the region corresponding to plus or minus one standard deviation.}
    \label{fig:energy_diff}
\end{figure}

\subsection*{The benchmark Gibbs-Boltzmann distribution}

In this paper, in order to test the $\Nf$ architecture on a prototypically hard-to-sample problem, we will consider the Edwards-Anderson 
spin glass model, described by the Hamiltonian in Eq.~\eqref{eq:spin_glass_H} for Ising spins, where we set $H_i=0$ for simplicity.
The couplings are chosen to be non-zero only for neighboring pairs $\< i,j \>$ on a two-dimensional \changes{square} lattice.
Non-zero couplings $J_{ij}$ are independent random variables, identically distributed according to either a normal or a Rademacher distribution. This model was also used as a benchmark in Refs.~\cite{mcnaughton_boosting_2020,ciarella_machine-learning-assisted_2023}.
All the results reported in the rest of the section are for a $16\times 16$ square lattice, hence with $N=256$ spins, which guarantees a large enough system to investigate correlations at low temperatures while keeping the problem tractable.
In the SI we also report some data for a $32\times 32$ lattice.

While this model has no finite-temperature spin glass transition,
sampling it via standard local MCMC becomes extremely hard for $\beta\gtrsim 1.5$~\cite{ciarella_machine-learning-assisted_2023}.
To this day, the best option for sampling this model at lower temperatures is parallel tempering (PT), which we used to construct a sample of $M=2^{16}$ equilibrium configurations for several instances of the model and at several temperatures.
Details are given in the Methods section.

The PT-generated training set is used to train the $\Nf$ architecture, independently for each instance and each temperature. Remember that the goal here is to test the expressivity of the architecture, which is why we train it in the best possible situation, i.e. by maximizing the likelihood of equilibrium configurations at each temperature.
Once the $\Nf$ model has been trained, we perform several comparisons that are reported in the following.

\subsection*{Energy}

We begin by comparing the mean energy of the configurations generated by the model with the mean energy of the configurations generated via parallel tempering, which is important because the energy plays a central role in the Metropolis reweighing scheme of Eq.~\eqref{eq:prob}. Fig.~\ref{fig:energy_diff} shows the relative energy difference,
as a function of inverse temperature $\beta$ and number of layers $\ell$, averaged over 10 different instances of the disorder.  First, it is clear that adding layers improves results remarkably. We stress that, since the activation functions of our network are linear, adding layers only helps in taking into account spins  that are further away. Second, we notice that the relative error found when using 6 layers or more is small for all temperatures considered. Finally, we notice that the behaviour of the error is non-monotonic, and indeed appears to decrease for low temperatures even for networks with few layers.

\begin{figure}[t]
    \centering
        \includegraphics[width=.45\textwidth]{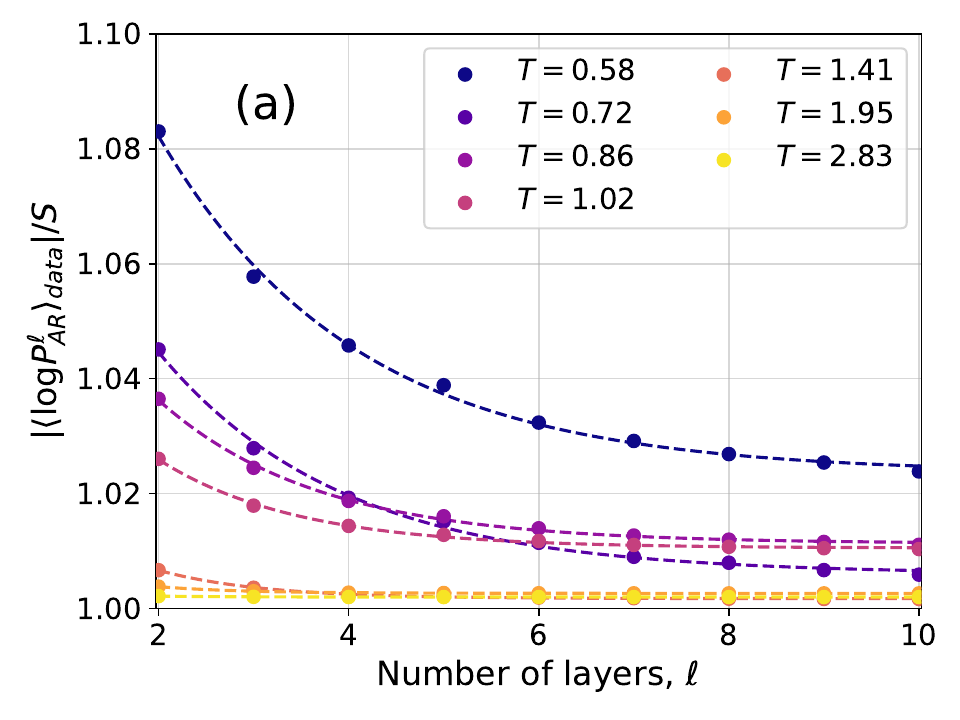}
        \label{fig:KL1}
        \includegraphics[width=.45\textwidth]{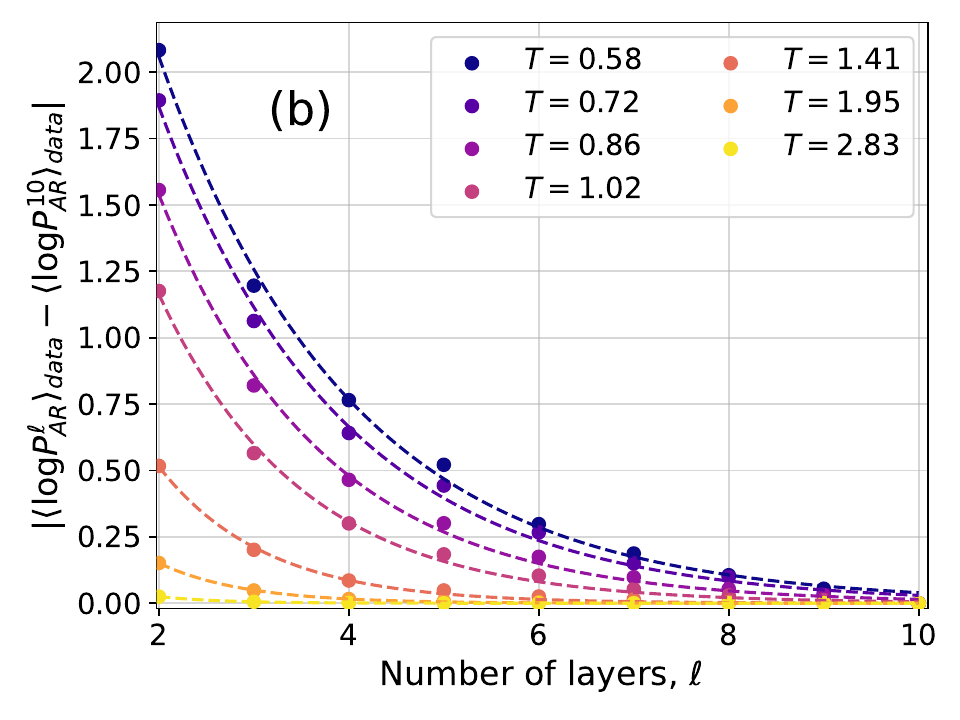}
        \label{fig:KL2}
    \caption{(a) Ratio between the cross-entropy $|\langle \log P^\ell_\text{NN} \rangle_\text{data}|$ and the Gibbs-Boltzmann entropy $S_\text{GB}$ for different values of the temperature $T$ and of the number of layers $\ell$. Both $|\langle \log P^\ell_\text{NN} \rangle_\text{data}|$ and $S_\text{GB}$ are averaged over 10 samples. Dashed lines are exponential fits in the form $Ae^{-\ell/\overline{\ell}} + C$. (b) Absolute difference between $\langle \log P^\ell_\text{NN} \rangle_\text{data}$ at various $\ell$ and at $\ell = 10$. Dashed lines are exponential fits in the form $Ae^{-\ell/\overline{\ell}}$.
    }
    \label{fig:KL}
\end{figure}

\begin{figure*}[t]
    \centering
    \begin{minipage}[b]{0.45\textwidth}
        \centering
        \includegraphics[width=\textwidth]
        {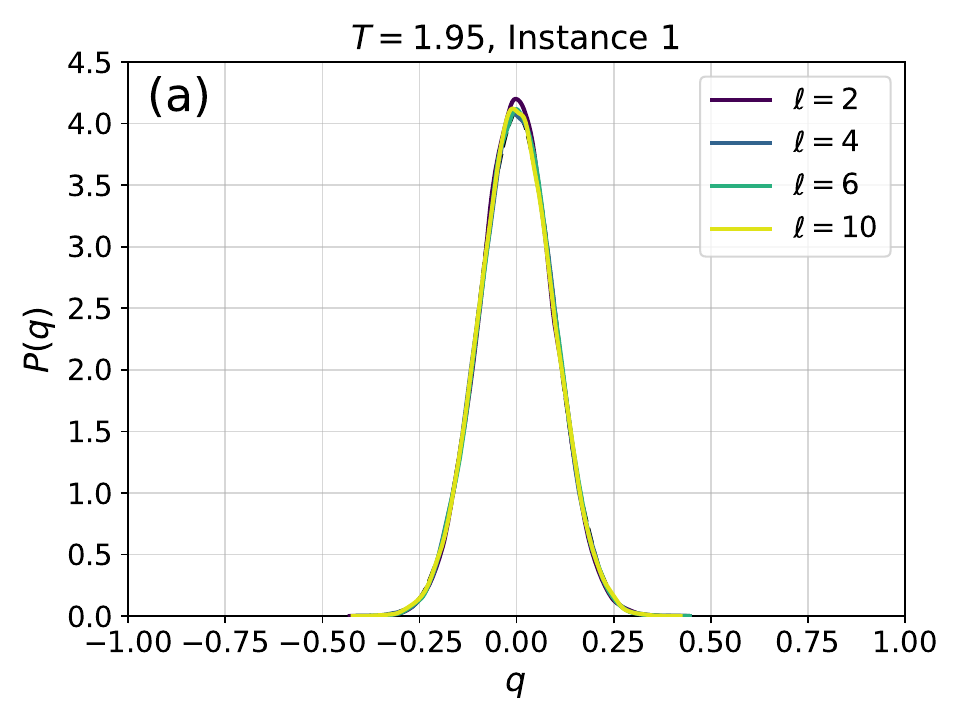}
    \end{minipage}
    \hfill
    \begin{minipage}[b]{0.45\textwidth}
        \centering
        \includegraphics[width=\textwidth]{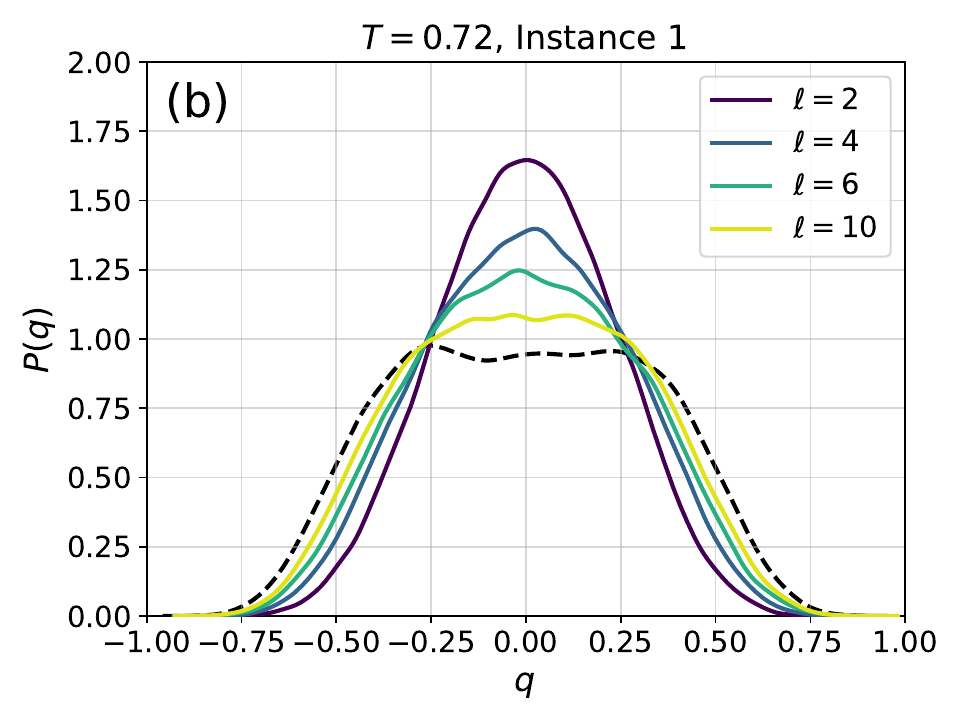}
        
    \end{minipage}
    \vskip\baselineskip
    \begin{minipage}[b]{0.45\textwidth}
        \centering
        \includegraphics[width=\textwidth]{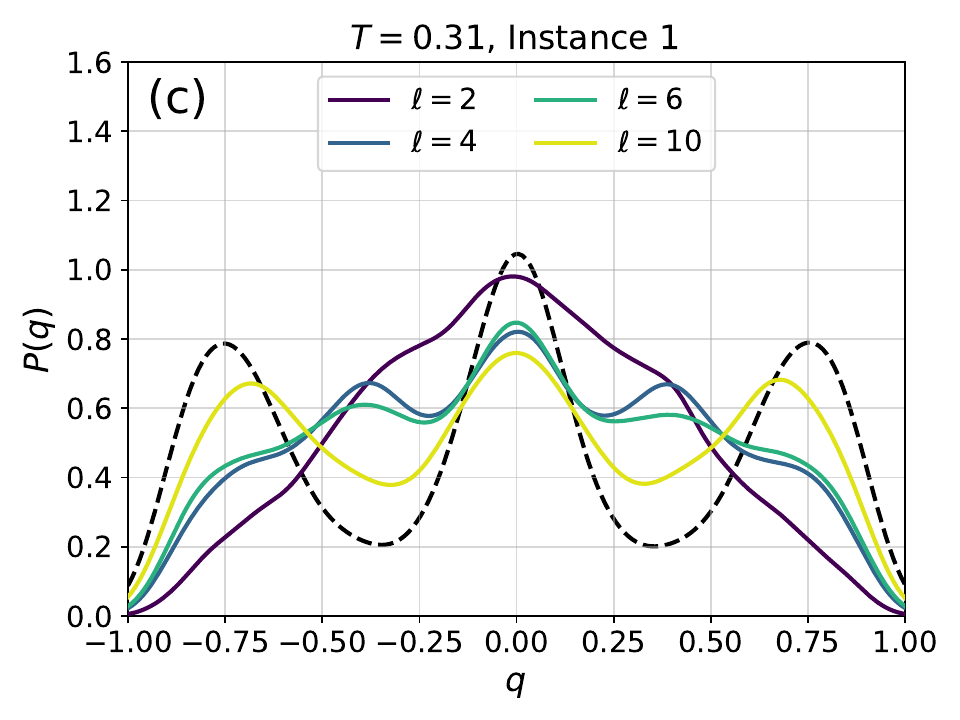}
        
    \end{minipage}
    \hfill
    \begin{minipage}[b]{0.45\textwidth}
        \centering
        \includegraphics[width=\textwidth]{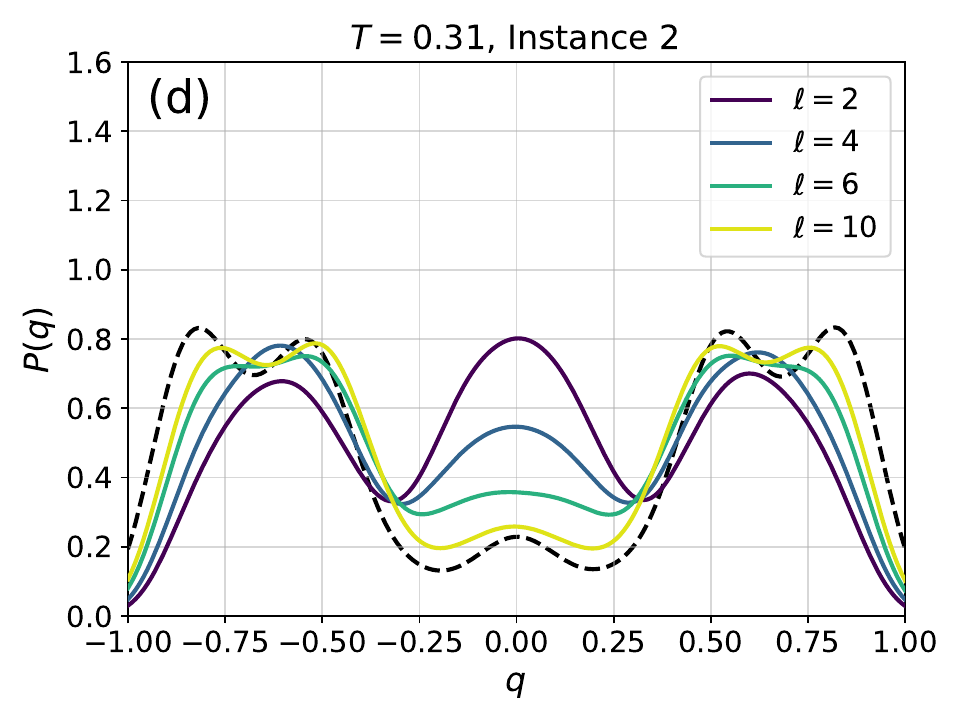}
        
    \end{minipage}
    \caption{Examples of the probability distribution of the overlap, $P(q)$, for different instances of the disorder and different temperatures. Solid color lines correspond to configurations generated using $\Nf$, while dashed black lines correspond to the $P(q)$ obtained from equilibrium configurations generated via PT. (a)~At high temperature, even a couple of layers are enough to reproduce well the distribution. (b)~At lower temperatures more layers are required, as expected because of the increase of the correlation length and of the complexity of the Gibbs-Boltzmann distribution.
    (c)~Even lower temperature for the same instance.
    (d)~A test on a more complex instance shows that $\Nf$ is capable to learn even non-trivial forms of $P(q)$. 
    }
    \label{fig:Pq}
\end{figure*}

\subsection*{Entropy and Kullback-Leibler divergence}

Next, we consider the Kullback-Leibler divergence between the learned probability distribution $P_\text{NN}$ and the target probability distribution, $D_{\text{KL}}(P_{\text{GB}} \parallel P_{\text{NN}})$. 
Indicating with $\<\, \cdot \, \>_\text{GB}$ the average with respect to $P_\text{GB}$, we have
\begin{equation}
    D_{\text{KL}}(P_{\text{GB}} \parallel P_{\text{NN}}) = \< \log P_\text{GB}\>_\text{GB} -\< \log P_\text{NN} \>_\text{GB} \ .
\end{equation}
The first term $\< \log P_\text{GB}\>_\text{GB}=-S_{\text{GB}}(T)$ is minus the entropy of the Gibbs-Boltzmann distribution and does not depend on $P_\text{NN}$, hence we focus on the cross-entropy $-\< \log P_\text{NN} \>_\text{GB}$ as a function of the number of layers for different temperatures. Minimizing this quantity corresponds to minimizing the KL divergence. In the case of perfect learning, i.e. $D_{\text{KL}} = 0$, we have $-\< \log P_\text{NN} \>_\text{GB}=S_{\text{GB}}(T)$. In Fig.~\ref{fig:KL} we compare the cross-entropy, estimated
as $-\< \log P_\text{NN} \>_\text{data}$ on the data generated by PT,
with $S_{\text{GB}}(T)$ computed using thermodynamic integration. We find a good match for large enough $\ell$, indicating an accurate training of the $\Nf$ model that does not suffer from mode collapse. 

In order to study more carefully how changing the number of layers affects the accuracy of the training, in Fig.~\ref{fig:KL} we have also shown the difference $|\langle \log P^\ell_{AR} \rangle_{data}- \langle \log P^{10}_{AR} \rangle_{data}|$ as a function of the number of layers $\ell$ for different temperatures. The plots are compatible with an exponential decay in the form $Ae^{-\ell/\overline{\ell}}$, with $\overline{\ell}$ thus being an estimate of the number of layers above which the $\Nf$ architecture becomes accurate.
We thus need to understand how $\overline{\ell}$ changes with temperature and system size. To do so, we need to introduce appropriate correlation functions, as we do next.

\begin{figure*}[t]
    \centering
    \begin{minipage}[b]{0.45\textwidth}
        \centering
        \includegraphics[width=\textwidth]{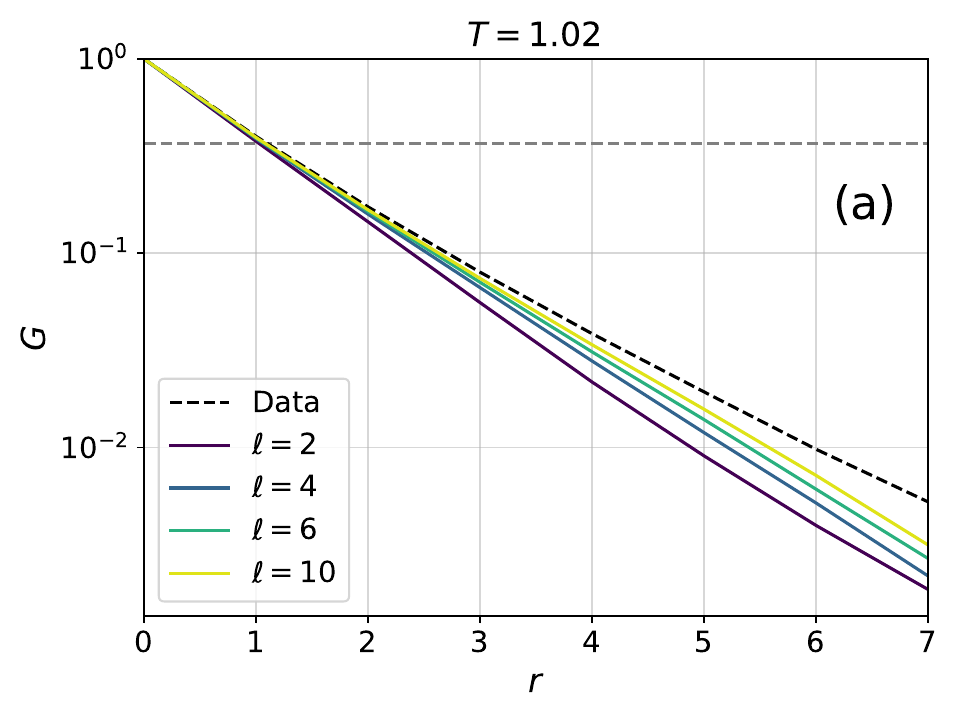}
    \end{minipage}
    \hfill
    \begin{minipage}[b]{0.45\textwidth}
        \centering
        \includegraphics[width=\textwidth]{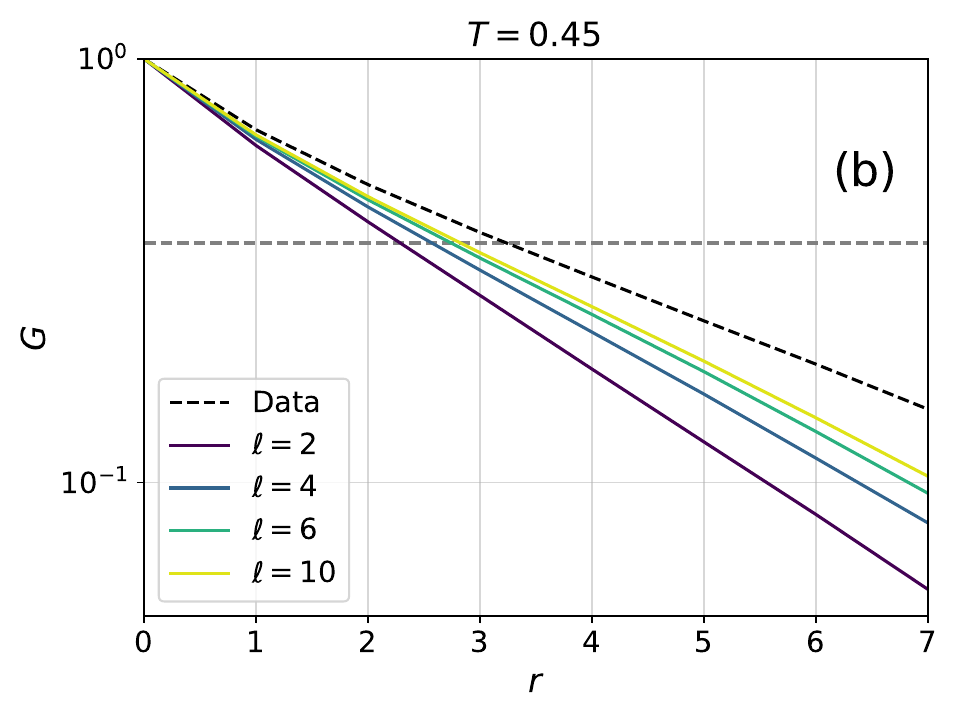}
    \end{minipage}
    \caption{Comparison between the spatial correlation functions obtained from $\Nf$ and those obtained from PT, at different high (a) and low (b) temperatures and for different number of layers. The grey dashed lines correspond to the value $1/e$ that defines the correlation length $\xi$. Data are averaged over 10 instances.}
    \label{fig:correlation_function}
\end{figure*}

\subsection*{Overlap distribution function}

In order to show that the $\Nf$ architecture is able to capture more complex features of the Gibbs-Boltzmann distribution, we have considered a more sensitive observable:  the probability distribution of the overlap, $P(q)$. The overlap $q$ between two configurations of spins $\ \boldsymbol{\sigma}^1 = \{ \sigma^1_1, \sigma^1_2, \dots \sigma^1_N \}$ and $\ \boldsymbol{\sigma}^2 = \{ \sigma^2_1, \sigma^2_2, \dots \sigma^2_N \}$ is a central quantity in spin glass physics, and is defined as:
\begin{equation}
    q_{12} = \frac{1}{N}\boldsymbol{\sigma}^1 \cdot \boldsymbol{\sigma}^2 = \frac{1}{N} \sum_{i = 1}^N \sigma^1_i \sigma^2_i.
\end{equation}
The histogram of $P(q)$ is obtained by considering many pairs of configurations, independently taken from either the PT-generated training set, or by generating them with the $\Nf$ architecture.
Examples for two different instances at several temperatures are shown in Fig.~\ref{fig:Pq}. At high temperatures, even very few layers are able to reproduce the distribution of the overlaps. When going to lower temperatures, and to more complex shapes of $P(q)$, however, the number of layers that is required to have a good approximation of $P(q)$ increases. This result is compatible with the intuitive idea that an increasing number of layers is required to reproduce a system at low temperatures, where the correlation length is large.  

As shown in the SI, the performance of $\Nf$ is on par with that of other algorithms with more parameters. Moreover, with a suitable increase of the number of layers, $\Nf$ is also able to achieve satisfactory results when system's size increases.

\begin{figure}[t] 
    \centering
\includegraphics[width=0.4
\textwidth]{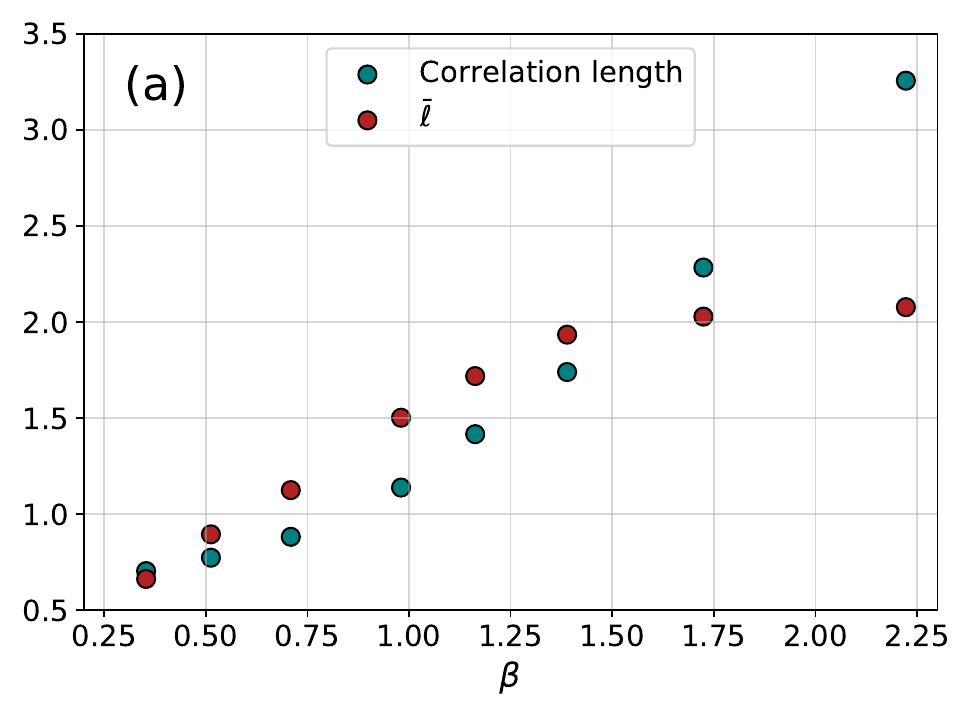}
    \includegraphics[width=.45\textwidth]{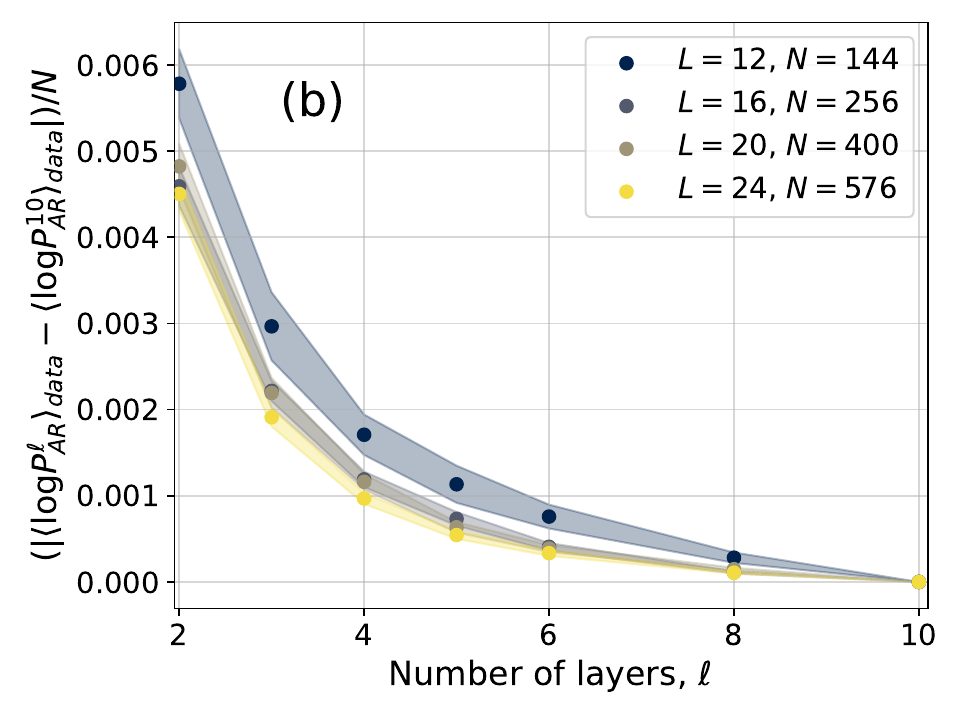}
        
    \caption{(\textit{a}) Comparison between the correlation length $\xi$ and the decay parameter $\overline{\ell}$ of the exponential fit as a function of the inverse temperature $\beta$. Both data sets are averaged over 10 instances. The data are limited to $T\geq 0.45$ because below that temperature the finite size of the sample prevents us to obtain a reliable proper estimate of the correlation length.
    (\textit{b})~Absolute difference between $\langle \log P^\ell_\text{NN} \rangle_\text{data}$ at various $\ell$ and at $\ell = 10$ for $T = 1.02$ and different system sizes. Data are averaged over ten different instances of the disorder and the colored areas identify regions corresponding to plus or minus one standard deviation. A version of this plot in linear-logarithmic scale is available in the SI.}\label{fig:xi_and_B}
\end{figure}

\subsection*{Spatial correlation function}

To better quantify spatial correlations, 
from the overlap field, a proper
correlation function for a spin glass system can be defined as~\cite{fernandez2016universal}
\begin{equation}
    G(r) = \frac{1}{\changes{\mathcal{N}(r)}} \sum_{i,j : d_{ij}=r} \overline{\langle q_i q_j \rangle} \ ,
\end{equation}
where $r=d_{ij}$ is the lattice distance of sites $i,j$, ${q_i = \sigma^1_i \sigma^2_i}$ is the local overlap,
and \changes{$\mathcal{N}(r) = \sum_{i,j: d_{ij}=r} 1$ is the number of pairs of sites at distance $r$}.
Note that in this way $G(0)=1$ by construction.
Results for $G(r)$ at various temperatures are shown in Fig.~\ref{fig:correlation_function}. It is clear that increasing the number of layers leads to a better agreement at lower temperatures. Interestingly, architectures with a lower numbers of layers produce a smaller correlation function at any distance, which shows that upon increasing the number of layers the $\Nf$ architecture is effectively learning the correlation of the Gibbs-Boltzmann measure.

To quantify this intuition, we extract from the exact $G(r)$ of the EA model a correlation length $\xi$, from the relation
$G(\xi)=1/e$. Then, we compare $\xi$ with $\overline{\ell}$, i.e. the number of layers above which the model becomes accurate, extracted from Fig.~\ref{fig:KL}.
These two quantities
are compared in Fig.~\ref{fig:xi_and_B} as a function of temperature, showing that they are roughly proportional. Actually $\overline{\ell}$ seems to grow even slower than $\xi$ at low temperatures.
In Fig.~\ref{fig:xi_and_B} we also reproduce the results of Fig.~\hyperref[fig:KL]{3b} for different sizes of the system. The curves coincide for large enough values of $N$, showing that the value of $\ell$ saturates to a system-independent value when increasing the system size. We are thus able to introduce a physical criterion to fix the optimal number of layers needed for the $\Nf$ architecture to perform well: $\ell$ must be of the order of the correlation length $\xi$ of the model at that temperature.

\changes{We recall that the $\Nf$ architecture requires $(c+1) \times \ell \times N$ parameters, where $c$ is the average connectivity of the graph ($c = 2d$ for a $d$-dimensional lattice), and $\mathcal{O}(\mathcal{B}(\ell)\times N)$ operations to generate a new configuration. 
We thus conclude that, for a model with a correlation length $\xi$ that remains finite for $N\to\infty$, we can choose $\ell$ proportional to $\xi$, hence achieving the desired linear scaling with system size both for the number of parameters and the number of operations.}

\section*{Discussion}\label{sec:conclusions}

In this work we have introduced a new, physics-inspired $\Nf$ architecture for the approximation of the Gibbs-Boltzmann probability distribution of a spin glass system of $N$ spins.
The $\Nf$ architecture is inspired by the previously introduced TwoBo~\cite{biazzo_sparse_2024}, but it implements explicitly the locality of the physical model via a Graph Neural Network structure. 

The $\Nf$ network possesses many desirable properties. (i)~The architecture has a physical interpretation, with the required number of layers being directly linked to the correlation length of the original data. (ii)~As a consequence, 
if the correlation length of the model is finite, the required number of parameters for a good fit scales linearly in the number of spins, thus greatly improving the scalability with respect to previous architectures. (iii)~The architecture is very general, and can be applied to lattices in all dimensions as well as arbitrary interaction graphs without losing the good scaling of the number of parameters, as long as the average connectivity remains finite. It can also be generalized to systems with more than two-body interactions by using a factor graph representation in the GNN layers.

As a benchmark study,
we have tested that the $\Nf$ architecture is powerful enough to reproduce key properties of the two-dimensional Edwards-Anderson spin glass model, such as the energy, the correlation function and the probability distribution of the overlap. 

Having tested that $\Nf$ achieves the best possible scaling with system size while being expressive enough,
the next step is to check whether it is possible to apply an iterative, self-consistent procedure to automatically train the network at lower and lower temperatures without the aid of training data generated with a different algorithm (here, parallel tempering). This would allow to set up a simulated tempering procedure with a good scaling in the number of variables, with applications not only in the simulation of disordered systems but in the solution of optimization problems as well. Whether setting up such a procedure is actually possible will be the focus of future work.

\section*{Methods}\label{sec:methods}

\subsection*{Details of the $\Nf$ architecture}

The Nearest Neighbors Neural Network ($\Nf$) architecture uses a GNN structure to respect the locality of the problem, as illustrated in Fig.~\ref{fig:N4scheme}. The input, given by the local fields computed from the already fixed spins, is passed through $\ell$ layers of weights $W_{ij}^k$. Each weight corresponds to a directed nearest neighbor link $i \leftarrow j$ (including $i = j$) and $k = 1, \dots, \ell$, therefore the network has a total number of parameters equal to $(c+1) \times \ell \times N$, i.e. linear in $N$ as long as the number of layers and the model connectivity remain finite. A showcase of the real wall-clock times required for the training of the network and for the generation of new configurations is shown in the SI.

The operations performed in order to compute ${\pi_i= P(\sigma_i=1 \mid \bs_{<i})}$ are the following:
\begin{enumerate}[leftmargin=0.4cm,label=\arabic*.]
    \item \textbf{initialize} the local fields from the input spins $\bs_{<i}$:
    \begin{equation}\nonumber
    h_l^{i,(0)} = \sum_{f(<i)} J_{lf} \sigma_f \ , \qquad \forall l : |i - l| \leq \ell \ .
    \end{equation}  
    \item \textbf{iterate} the GNN layers
    $\forall k = 1,\cdots, \ell-1$:
    \begin{equation}\nonumber
    h_l^{i,(k)} = \phi^k \left[ \beta W_{ll}^k h_l^{i,(k-1)} + \sum_{j \in \partial l} \beta J_{lj} W_{lj}^k h_j^{i,(k-1)} \right] + h_l^{i,(0)} \ ,
    \end{equation}
    where the sum over $j\in \partial l$ runs over the neighbors of site $l$. The fields are computed $\forall l : |i - l| \leq \ell-k$, see Fig.~\ref{fig:N4scheme}.
    \item \textbf{compute} the field on site $i$ at the final layer, \[ h_i^{i,(\ell)} = \phi^\ell \left[ \beta W_{ii}^\ell h_i^{i,(\ell-1)} + \sum_{j \in \partial i} \beta J_{ij} W_{ij}^\ell h_j^{i,(\ell-1)} \right] \ . \]
    \item \textbf{output} \( \pi_i =P(\sigma_i = 1 | \bs_{<i}) = S[2h_i^{i,(\ell)}] \).
\end{enumerate}
Here $S(x) = 1/(1+e^{-x})$ is the sigmoid function and $\phi^k(x)$ are layer-specific functions that can be used to introduce non-linearities in the system. The fields $h_l^{i,(0)}=\xi_{il}$ in the notation of the TwoBo architecture. Differently from TwoBo, however, $\Nf$ uses all the $h_l^{i,(0)}$, instead of the ones corresponding to spins that have not yet been fixed. 

In practice, we have observed that the network with identity functions, i.e. $\phi^k(x) = x$ for all $k$, is sufficiently expressive, and using non-linearities does not seem to improve the results. Moreover, linear functions have the advantage of being easily explainable and faster, therefore we have chosen to use them in this work.

\subsection*{Spin update order}

Autoregressive models such as $\Nf$ are based on an ordering of spins to implement Eq.~\eqref{eq:factorization}.
One can choose a simple raster scan (i.e., from left to right on one row and updating each row from top to bottom sequentially) or any other scheme of choice, see e.g. Refs.~\cite{trinquier2021efficient,teoh2024autoregressive} for a discussion. 
In this work, we have adopted a spiral scheme, in which the spins at the edges of the system are updated first, and then one moves in spirals towards the central spins. As shown in the SI, it appears as $\Nf$ performs better, for equal training conditions, when the spins are updated following a spiral scheme with respect to, for instance, a raster scheme.

\subsection*{Building the training data}

We use a training set of $M=2^{16}$ configurations obtained at several temperatures using a Parallel Tempering algorithm for different instances of the couplings $J_{ij}$, and for a 16x16 two-dimensional square lattice with open boundary conditions. Data where gathered in runs of $2^{27}$ steps by considering $2^{26}$ steps for thermalization and then taking one configuration each $2^{10}$ steps. Temperature swap moves, in which one swaps two configurations at nearby temperature, were attempted once every 32 normal Metropolis sweeps. 
The choice of $M=2^{16}$ was motivated based on previous studies of the same model~\cite{ciarella_machine-learning-assisted_2023}.

\subsection*{Model training}

The training has been carried out maximizing the likelihood of the data, a procedure which is equivalent to minimizing the KL divergence $D_{\text{KL}}(P_{\text{GB}} \parallel P_{\text{NN}})$, with $P_{\text{GB}}$ estimated by the training set.
 We used an ADAM optimizer with batch size 128 for 80 epochs and a learning rate 0.001 with exponential decay in time. Training is stopped if the loss has not improved in the last 10 steps, and the lowest-loss model is considered.

\acknowledgments

We especially thank Indaco Biazzo and Giuseppe Carleo for several important discussions on their TwoBo architecture that motivated and inspired the present study. We also thank Giulio Biroli, Patrick Charbonneau, Simone Ciarella, Marylou Gabri\'e, Leonardo Galliano, Jeanne Trinquier, Martin Weigt and Lenka Zdeborov\'a for useful discussions. \changes{We acknowledge  support from the computational infrastructure DARIAH.IT, PON Project code PIR01\_00022, National Research Council of Italy. The research has received financial support from the “National Centre for HPC, Big Data and Quantum Computing - HPC”, Project CN\_00000013, CUP B83C22002940006, NRP Mission 4 Component 2 Investment 1.5,  Funded by the European Union - NextGenerationEU}. Author LMDB acknowledges funding from the Bando Ricerca Scientifica 2024 - \textit{Avvio alla Ricerca} (D.R. No. 1179/2024) of Sapienza Università di Roma, project B83C24005280001 – MaLeDiSSi.

\bibliographystyle{unsrt}
\bibliography{references}

\begin{thebibliography}{10}

\bibitem{wolff_collective_1989}
Ulli Wolff.
\newblock Collective monte carlo updating for spin systems.
\newblock {\em Physical Review Letters}, 62(4):361--364, 1989.
\newblock Publisher: American Physical Society.

\bibitem{Ninarello2017}
Andrea Ninarello, Ludovic Berthier, and Daniele Coslovich.
\newblock Models and algorithms for the next generation of glass transition studies.
\newblock {\em Phys. Rev. X}, 7:021039, Jun 2017.

\bibitem{Marinari}
E.~Marinari and G.~Parisi.
\newblock Simulated tempering: A new monte carlo scheme.
\newblock {\em Europhysics Letters}, 19(6):451, jul 1992.

\bibitem{houdayer_cluster_2001}
J.~Houdayer.
\newblock A cluster monte carlo algorithm for 2-dimensional spin glasses.
\newblock {\em The European Physical Journal B - Condensed Matter and Complex Systems}, 22(4):479--484, 2001.

\bibitem{carleo_solving_2017}
Giuseppe Carleo and Matthias Troyer.
\newblock Solving the quantum many-body problem with artificial neural networks.
\newblock {\em Science}, 355(6325):602--606, 2017.
\newblock Publisher: American Association for the Advancement of Science.

\bibitem{tamagnone2024coarse}
Samuel Tamagnone, Alessandro Laio, and Marylou Gabri{\'e}.
\newblock Coarse grained molecular dynamics with normalizing flows.
\newblock {\em arXiv preprint arXiv:2406.01524}, 2024.

\bibitem{jung2023roadmap}
Gerhard Jung, Rinske~M Alkemade, Victor Bapst, Daniele Coslovich, Laura Filion, Fran{\c{c}}ois~P Landes, Andrea Liu, Francesco~Saverio Pezzicoli, Hayato Shiba, Giovanni Volpe, et~al.
\newblock Roadmap on machine learning glassy liquids.
\newblock {\em arXiv preprint arXiv:2311.14752}, 2023.

\bibitem{galliano2024policy}
Leonardo Galliano, Riccardo Rende, and Daniele Coslovich.
\newblock Policy-guided monte carlo on general state spaces: Application to glass-forming mixtures.
\newblock {\em arXiv preprint arXiv:2407.03275}, 2024.

\bibitem{ciarella2023finding}
Simone Ciarella, Dmytro Khomenko, Ludovic Berthier, Felix~C Mocanu, David~R Reichman, Camille Scalliet, and Francesco Zamponi.
\newblock Finding defects in glasses through machine learning.
\newblock {\em Nature Communications}, 14(1):4229, 2023.

\bibitem{gabrie_adaptive_2022}
Marylou Gabrié, Grant~M. Rotskoff, and Eric Vanden-Eijnden.
\newblock Adaptive monte carlo augmented with normalizing flows.
\newblock {\em Proceedings of the National Academy of Sciences}, 119(10):e2109420119, 2022.
\newblock Publisher: Proceedings of the National Academy of Sciences.

\bibitem{wu_solving_2019}
Dian Wu, Lei Wang, and Pan Zhang.
\newblock Solving statistical mechanics using variational autoregressive networks.
\newblock {\em Physical Review Letters}, 122(8):080602, 2019.
\newblock Publisher: American Physical Society.

\bibitem{ciarella_machine-learning-assisted_2023}
Simone Ciarella, Jeanne Trinquier, Martin Weigt, and Francesco Zamponi.
\newblock Machine-learning-assisted monte carlo fails at sampling computationally hard problems.
\newblock {\em Machine Learning: Science and Technology}, 4(1):010501, 2023.

\bibitem{mcnaughton_boosting_2020}
B.~{McNaughton}, M.~V. Milošević, A.~Perali, and S.~Pilati.
\newblock Boosting monte carlo simulations of spin glasses using autoregressive neural networks.
\newblock {\em Physical Review E}, 101(5):053312, 2020.
\newblock Publisher: American Physical Society.

\bibitem{trinquier2021efficient}
Jeanne Trinquier, Guido Uguzzoni, Andrea Pagnani, Francesco Zamponi, and Martin Weigt.
\newblock Efficient generative modeling of protein sequences using simple autoregressive models.
\newblock {\em Nature communications}, 12(1):5800, 2021.

\bibitem{kohler_equivariant_2020}
Jonas Köhler, Leon Klein, and Frank Noé.
\newblock Equivariant flows: Exact likelihood generative learning for symmetric densities.

\bibitem{dibak_temperature_2022}
Manuel Dibak, Leon Klein, Andreas Krämer, and Frank Noé.
\newblock Temperature steerable flows and boltzmann generators.
\newblock {\em Nature Machine Intelligence}, 4(4):L042005, 2022.
\newblock Publisher: American Physical Society.

\bibitem{invernizzi2022skipping}
Michele Invernizzi, Andreas Kr\"amer, Cecilia Clementi, and Frank No{\'e}.
\newblock Skipping the replica exchange ladder with normalizing flows.
\newblock {\em The Journal of Physical Chemistry Letters}, 13(50):11643--11649, 2022.

\bibitem{chen2024diffusive}
Wenlin Chen, Mingtian Zhang, Brooks Paige, Jos{\'e}~Miguel Hern{\'a}ndez-Lobato, and David Barber.
\newblock Diffusive gibbs sampling.
\newblock {\em arXiv preprint arXiv:2402.03008}, 2024.

\bibitem{ghio2023sampling}
Davide Ghio, Yatin Dandi, Florent Krzakala, and Lenka Zdeborov{\'a}.
\newblock Sampling with flows, diffusion and autoregressive neural networks: A spin-glass perspective.
\newblock {\em arXiv preprint arXiv:2308.14085}, 2023.

\bibitem{kilgour2022inside}
Michael Kilgour and Lena Simine.
\newblock Inside the black box: A physical basis for the effectiveness of deep generative models of amorphous materials.
\newblock {\em Journal of Computational Physics}, 452:110885, 2022.

\bibitem{biazzo_sparse_2024}
Indaco Biazzo, Dian Wu, and Giuseppe Carleo.
\newblock Sparse autoregressive neural networks for classical spin systems.

\bibitem{schuetz_combinatorial_2022}
Martin J.~A. Schuetz, J.~Kyle Brubaker, and Helmut~G. Katzgraber.
\newblock Combinatorial optimization with physics-inspired graph neural networks.
\newblock {\em Nature Machine Intelligence}, 4(4):367--377, 2022.
\newblock Publisher: Nature Publishing Group.

\bibitem{schuetz_graph_2022}
Martin J.~A. Schuetz, J.~Kyle Brubaker, Zhihuai Zhu, and Helmut~G. Katzgraber.
\newblock Graph coloring with physics-inspired graph neural networks.
\newblock {\em Physical Review Research}, 4(4):043131, 2022.
\newblock Publisher: American Physical Society.

\bibitem{fan_searching_2023}
Changjun Fan, Mutian Shen, Zohar Nussinov, Zhong Liu, Yizhou Sun, and Yang-Yu Liu.
\newblock Searching for spin glass ground states through deep reinforcement learning.
\newblock {\em Nature Communications}, 14(1):725, February 2023.
\newblock Publisher: Nature Publishing Group.

\bibitem{pugacheva2024enhancing}
Daria Pugacheva, Andrei Ermakov, Igor Lyskov, Ilya Makarov, and Yuriy Zotov.
\newblock Enhancing gnns performance on combinatorial optimization by recurrent feature update.
\newblock {\em arXiv:2407.16468}, 2024.

\bibitem{colantonio2024efficient}
Lorenzo Colantonio, Andrea Cacioppo, Federico Scarpati, and Stefano Giagu.
\newblock Efficient graph coloring with neural networks: A physics-inspired approach for large graphs.
\newblock {\em arXiv:2408.01503}, 2024.

\bibitem{angelini_modern_2022}
Maria~Chiara Angelini and Federico Ricci-Tersenghi.
\newblock Modern graph neural networks do worse than classical greedy algorithms in solving combinatorial optimization problems like maximum independent set.
\newblock {\em Nature Machine Intelligence}, 5(1):29--31, December 2022.
\newblock arXiv:2206.13211 [cond-mat].

\bibitem{boettcher2023inability}
Stefan Boettcher.
\newblock Inability of a graph neural network heuristic to outperform greedy algorithms in solving combinatorial optimization problems.
\newblock {\em Nature Machine Intelligence}, 5(1):24--25, 2023.

\bibitem{boettcher2024deep}
Stefan Boettcher.
\newblock Deep reinforced learning heuristic tested on spin-glass ground states: The larger picture.
\newblock {\em Nature Communications}, 14(1):5658, 2023.

\bibitem{fan2023reply}
Changjun Fan, Mutian Shen, Zohar Nussinov, Zhong Liu, Yizhou Sun, and Yang-Yu Liu.
\newblock Reply to: Deep reinforced learning heuristic tested on spin-glass ground states: The larger picture.
\newblock {\em Nature communications}, 14(1):5659, 2023.

\bibitem{germain2015made}
Mathieu Germain, Karol Gregor, Iain Murray, and Hugo Larochelle.
\newblock Made: Masked autoencoder for distribution estimation.
\newblock In {\em International conference on machine learning}, pages 881--889. PMLR, 2015.

\bibitem{van2016conditional}
Aaron Van~den Oord, Nal Kalchbrenner, Lasse Espeholt, Oriol Vinyals, Alex Graves, et~al.
\newblock Conditional image generation with pixelcnn decoders.
\newblock {\em Advances in neural information processing systems}, 29, 2016.

\bibitem{madanchi2024simulations}
Ata Madanchi, Michael Kilgour, Frederik Zysk, Thomas~D K{\"u}hne, and Lena Simine.
\newblock Simulations of disordered matter in 3d with the morphological autoregressive protocol (map) and convolutional neural networks.
\newblock {\em The Journal of Chemical Physics}, 160(2), 2024.

\bibitem{biazzo2023autoregressive}
Indaco Biazzo.
\newblock The autoregressive neural network architecture of the boltzmann distribution of pairwise interacting spins systems.
\newblock {\em Communications Physics}, 6(1):296, 2023.

\bibitem{fernandez2016universal}
LA~Fernandez, E~Marinari, V~Martin-Mayor, Giorgio Parisi, and JJ~Ruiz-Lorenzo.
\newblock Universal critical behavior of the two-dimensional ising spin glass.
\newblock {\em Physical Review B}, 94(2):024402, 2016.

\bibitem{teoh2024autoregressive}
Yi~Hong Teoh and Roger~G Melko.
\newblock Autoregressive model path dependence near ising criticality.
\newblock {\em arXiv:2408.15715}, 2024.

\end{thebibliography}

\pagebreak

\onecolumngrid

\appendix

\renewcommand{\thesection}{S\arabic{section}} 
\renewcommand{\appendixname}{}
\renewcommand{\thefigure}{S\arabic{figure}}
\setcounter{figure}{0}

\pagebreak

\pagenumbering{gobble}

\section{Computational details}

We now introduce a easy way to generate efficiently data using the 4N architecture. As long as the activation functions $\phi^k$ of the different layers of $\Nf$ are linear, all the intermediate message passing layers can be substituted with a single matrix multiplication. Indeed, let us consider, as in the main text, identity activation functions. In this case, we can rewrite the matrix implementing the $t-$th layer, $\underline{\underline{A}}^t$, as $\underline{\underline{A}}^t = \beta(\underline{\underline{J}} + \underline{\underline{I}})\odot \underline{\underline{W}}^t$, where $\beta$ is the inverse temperature, $\underline{\underline{J}}$ is the coupling matrix, $\underline{\underline{I}}$ is the identity matrix, and $\underline{\underline{W}}^t$ it the $t-$th layer weight matrix. Here,  $\odot$ is the elementwise multiplication. Now we observe that (in $\Nf$ we put skip connections after all but the last layer)
\begin{align}
\mathbf{h}^t = \underline{\underline{A}}^t \mathbf{h}^{t-1} = \underline{\underline{A}}^t (\underline{\underline{A}}^{t-1} \mathbf{h}^{t-2} +h^0) = \dots = (\underline{\underline{A}}^t \underline{\underline{A}}^{t-1}\dots \underline{\underline{A}}^1 + \underline{\underline{A}}^t \underline{\underline{A}}^{t-1}\dots \underline{\underline{A}}^2 + \dots + \underline{\underline{A}}^t \underline{\underline{A}}^{t-1} + \underline{\underline{A}}^t) \mathbf{h}^0.
\end{align}
Now, remembering that, if the input configuration $\mathbf{s}$ is properly masked, we have $\mathbf{h}^0 = \underline{\underline{J}}\mathbf{s}$, it follows that
\begin{align}
\mathbf{h}^t = (\underline{\underline{A}}^t \underline{\underline{A}}^{t-1}\dots \underline{\underline{A}}^1 + \underline{\underline{A}}^t \underline{\underline{A}}^{t-1}\dots \underline{\underline{A}}^2 + \dots + \underline{\underline{A}}^t \underline{\underline{A}}^{t-1} + \underline{\underline{A}}^t) \underline{\underline{J}}\mathbf{s},
\end{align}
which can be rewritten in matrix form as $\mathbf{h}^t = \underline{\underline{M}}\mathbf{s}$ by defining
\begin{align}
    \underline{\underline{M}} = (\underline{\underline{A}}^t \underline{\underline{A}}^{t-1}\dots \underline{\underline{A}}^1 + \underline{\underline{A}}^t \underline{\underline{A}}^{t-1}\dots \underline{\underline{A}}^2 + \dots + \underline{\underline{A}}^t \underline{\underline{A}}^{t-1} + \underline{\underline{A}}^t) \underline{\underline{J}},
\end{align}
and the final output is obtained simply by taking the elementwise sigmoidal of $\mathbf{h}^t$. Notice that the $i$-th row of the matrix $\underline{\underline{M}}$ has zeroes in all columns corresponding to spins at distance larger than $\ell + 1$ from the spin $i$, that is, the number of non-zero elements is constant with system size as long as $\ell$ is kept fixed. Therefore, generation can simply be implemented as a series of vector products, each considering only the nonzero elements of the matrix $\underline{\underline{M}}$. This allows one to compute the $i$-th element of vector $\mathbf{h}^t$ in a time $\mathcal{O}(1)$ in system size, therefore keeping generation procedures $\mathcal{O}(N)$. Moreover, the training time scales as $\mathcal{O}(N)$.

We point out that we can train directly the whole matrix $\underline{\underline{M}}$, with the additional constraint of putting to zero the elements corresponding to spins at distance larger than $\ell+1$. Conceptually, this approach can be interpreted as a reduced MADE, in which the weight matrix is not only triangular but has additional zeros. We will refer to this approach as $\ell$-MADE in the following. This method is easier to implement and can be faster in the training and generation processes, but has the disadvantage or requiring more parameters than the standard $\Nf$ approach, \changes{e.g. $\mathcal{O}(\ell^d N)$ for a $d-$dimensional lattice} against the $(c+1)\ell N$ of $\Nf$. Moreover, it loses the physical interpretation of $\Nf$, as it can no longer be interpreted as a series of fields propagating on the lattice.

\begin{figure*}[ht]
    \centering
    \begin{minipage}[b]{0.45\textwidth}
        \centering
        \includegraphics[width=\textwidth]{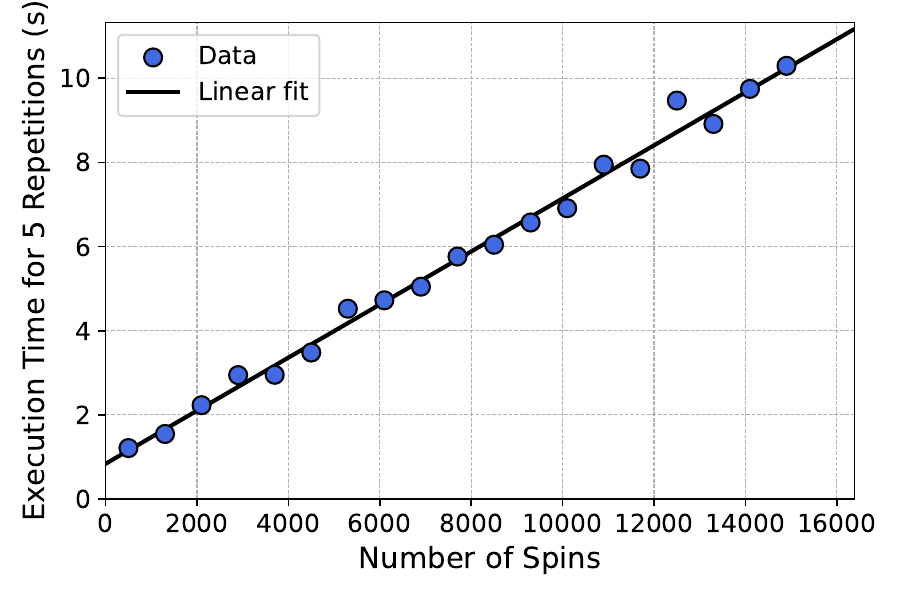}
        
    \end{minipage}
    \hfill
    \begin{minipage}[b]{0.45\textwidth}
        \centering
        \includegraphics[width=\textwidth]{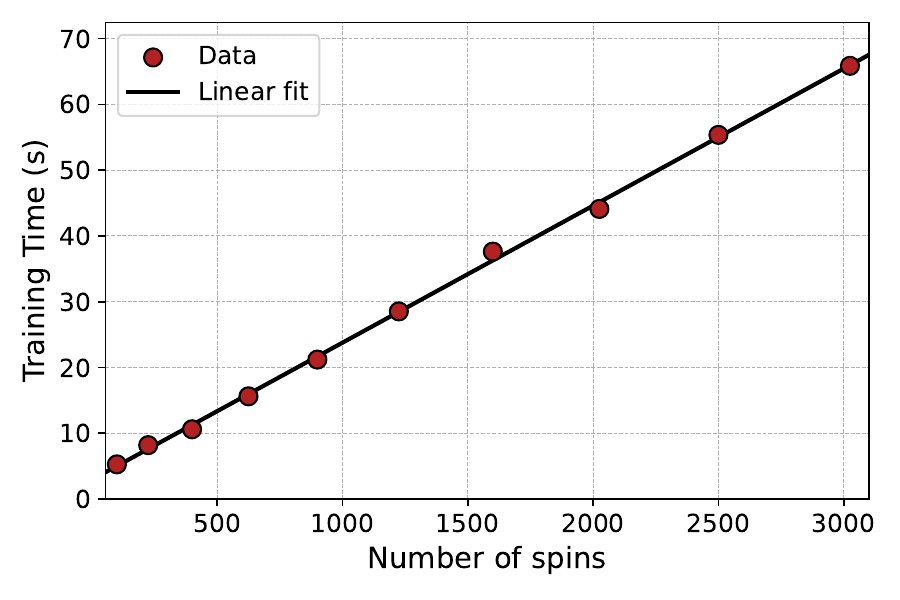}
        
    \end{minipage}
    \caption{Scatter plots of the generation (\textit{left}) and training (\textit{right}) elapsed real times at at different system sizes for $\ell$-MADE. Both training and generations have been performed using dummy data (and therefore $N$ needs not to be the square of an integer). For generation, we generate 10000 configurations of $N$ five times. For training, we train a model with $\ell = 2$ for 40 epochs using an Adam optimizer once per each value of $N$. In both cases, elapsed times scale linearly with the system's size, as further evidenced by the linear fits (black lines).
    }
    \label{fig:SIwallclock}
\end{figure*}

\begin{figure*}[ht]
    \centering
    \begin{minipage}[b]{0.45\textwidth}
        \centering
        \includegraphics[width=\textwidth]{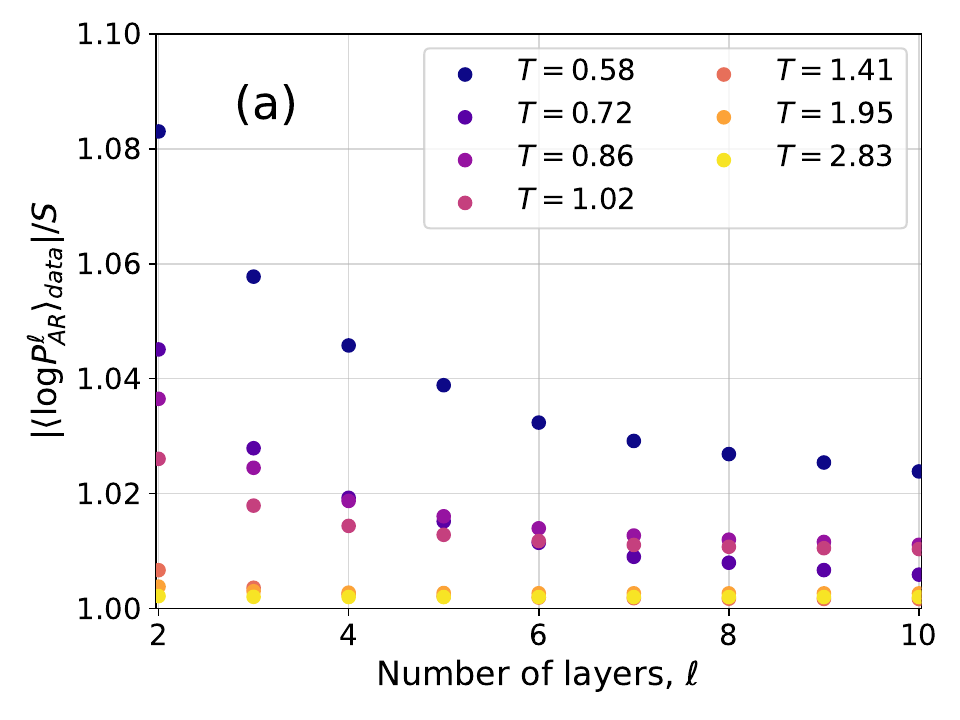}
        
    \end{minipage}
    \hfill
    \begin{minipage}[b]{0.45\textwidth}
        \centering
        \includegraphics[width=\textwidth]{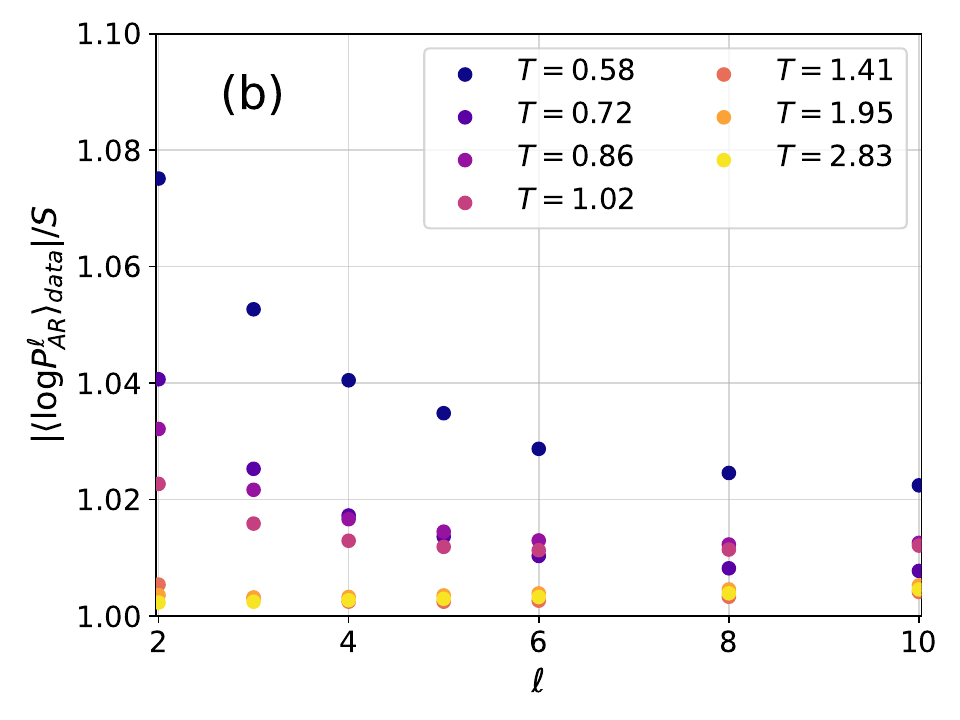}
        
    \end{minipage}
    \caption{Ratio between the cross-entropy $|\langle \log P^\ell_\text{NN} \rangle_\text{data}|$ and the Gibbs-Boltzmann entropy $S_\text{GB}$ for different values of the temperature $T$ and of the parameter $\ell$ for $\Nf$ (\textit{a, left}) and $\ell$-MADE (\textit{b, right}). Both $|\langle \log P^\ell_\text{NN} \rangle_\text{data}|$ and $S_\text{GB}$ are averaged over 10 samples.
    }
    \label{fig:ellMADE_comparison}
\end{figure*}

Wall-clock generation and training times are shown in Fig.~\ref{fig:SIwallclock}. Comparison between Fig.~\hyperref[fig:KL]{3b} obtained using $\Nf$ and the analogue obtained using $\ell$-MADE is shown in Fig.~\ref{fig:ellMADE_comparison}.

%
%
%
%

\clearpage

\section{Comparison between different architectures}

\vspace{-60em}

\begin{figure*}[!th]
    \centering
    \begin{minipage}[b]{0.45\textwidth}
        \centering
        \includegraphics[width=\textwidth]{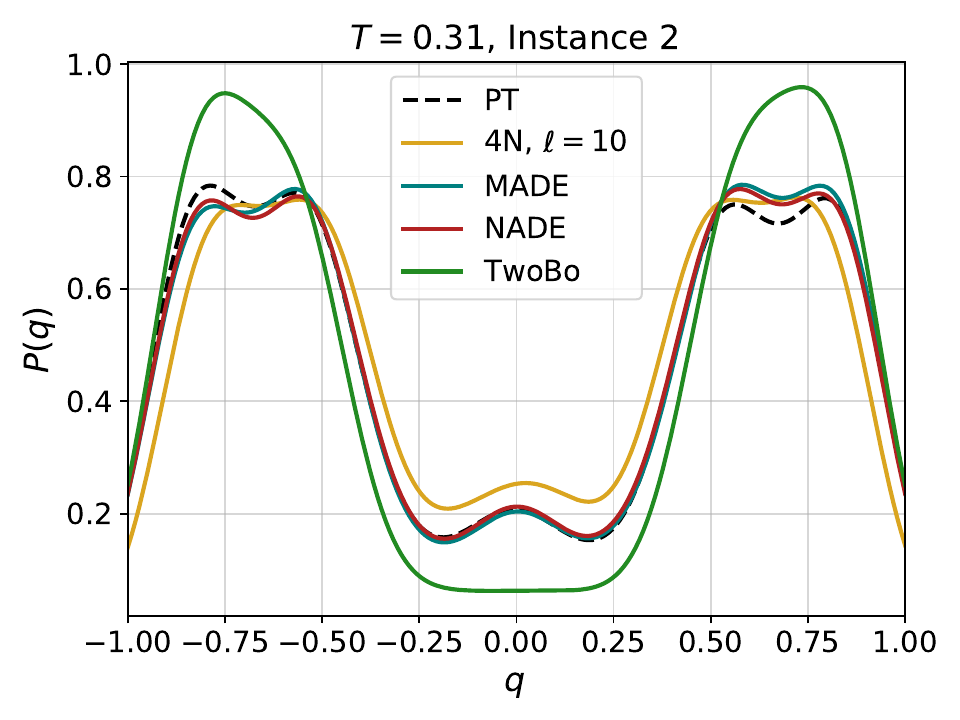}
        
    \end{minipage}
    \hfill
    \begin{minipage}[b]{0.45\textwidth}
        \centering
        \includegraphics[width=\textwidth]{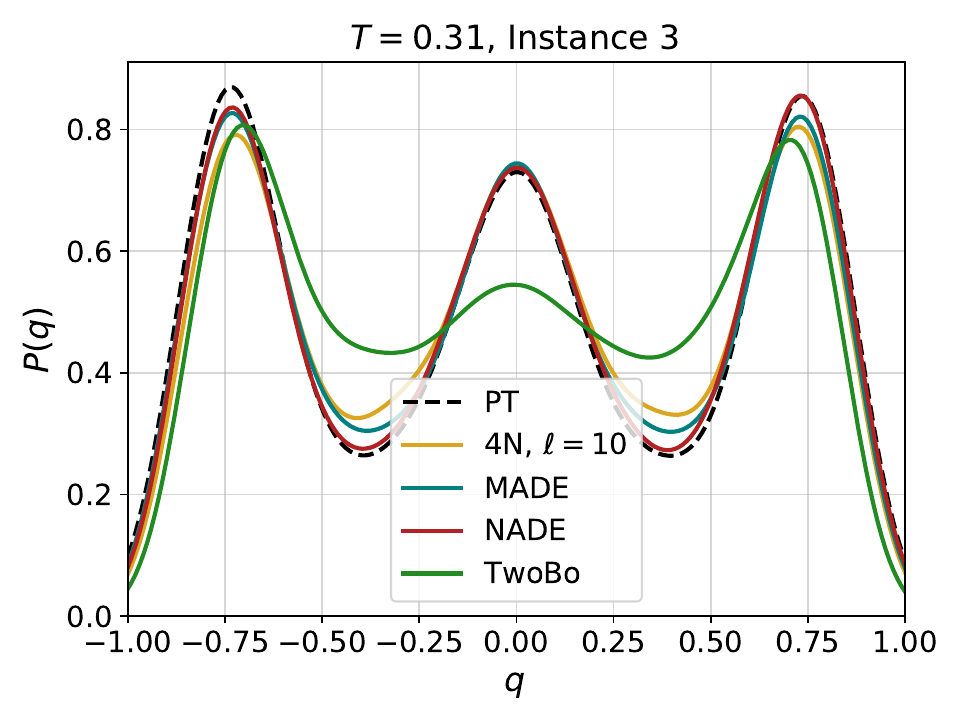}
        
    \end{minipage}
    %
    %
    \caption{Comparison of the probability distribution of the overlap obtained using different architectures: $\Nf$ with $\ell = 10$ layers (12160 parameters), shallow MADE (32640 parameters), NADE with $N_h = 64$ (32768 parameters) and TwoBo (4111 parameters). The black dashed line is obtained using data gathered via Parallel Tempering (PT), while solid color lines are obtained using the different architectures. \changes{From a quick qualitative analysis} $\Nf$ performs comparably to MADE and NADE despite having less parameters. Moreover, it also performs \changes{better than} TwoBo. Despite the latter having fewer parameters in this $N = 256$ case, $\Nf$ has a better scaling when increasing the system size, $\mathcal{O}(N)$ compared to $\mathcal{O}(N^\frac{3}{2})$, as pointed out in the main text. The TwoBo implementation comes from the GitHub of the authors (https://github.com/cqsl/sparse-twobody-arnn) and was slightly modified to support supervised learning (which could be improved by a different choice of the hyperparameters used for training).}
    \label{fig:PqSI2}
\end{figure*}

\pagebreak

\section{Comparison between different update sequences}

\vspace{-60em}

\begin{figure*}[!th]
    \centering
    \begin{minipage}[b]{0.45\textwidth}
        \centering
        \includegraphics[width=\textwidth]{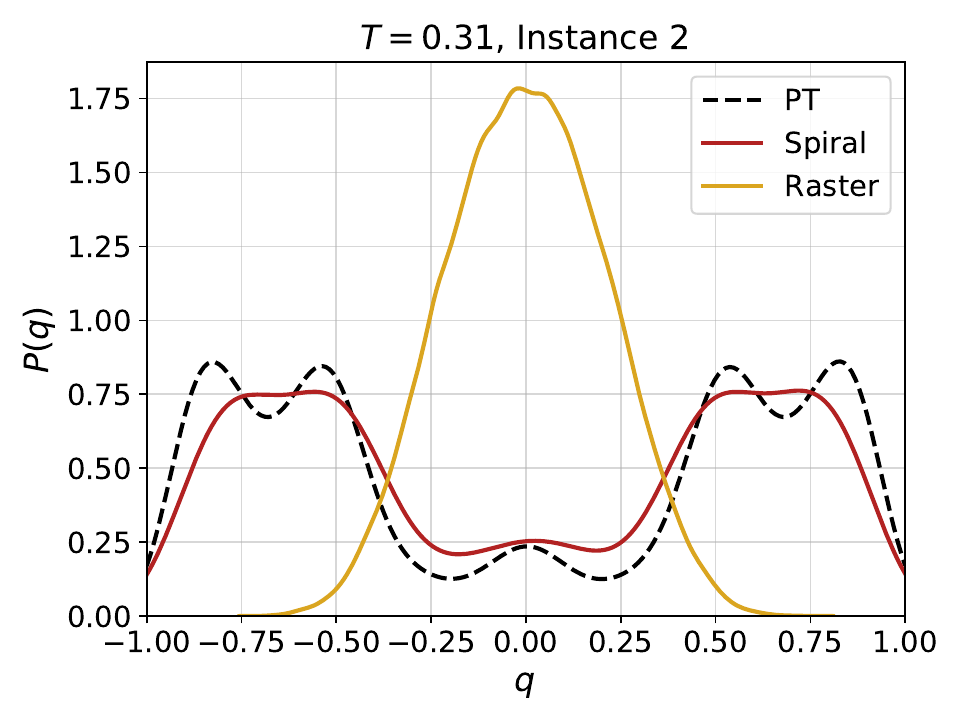}
        
    \end{minipage}
    \hfill
    \begin{minipage}[b]{0.45\textwidth}
        \centering
        \includegraphics[width=\textwidth]{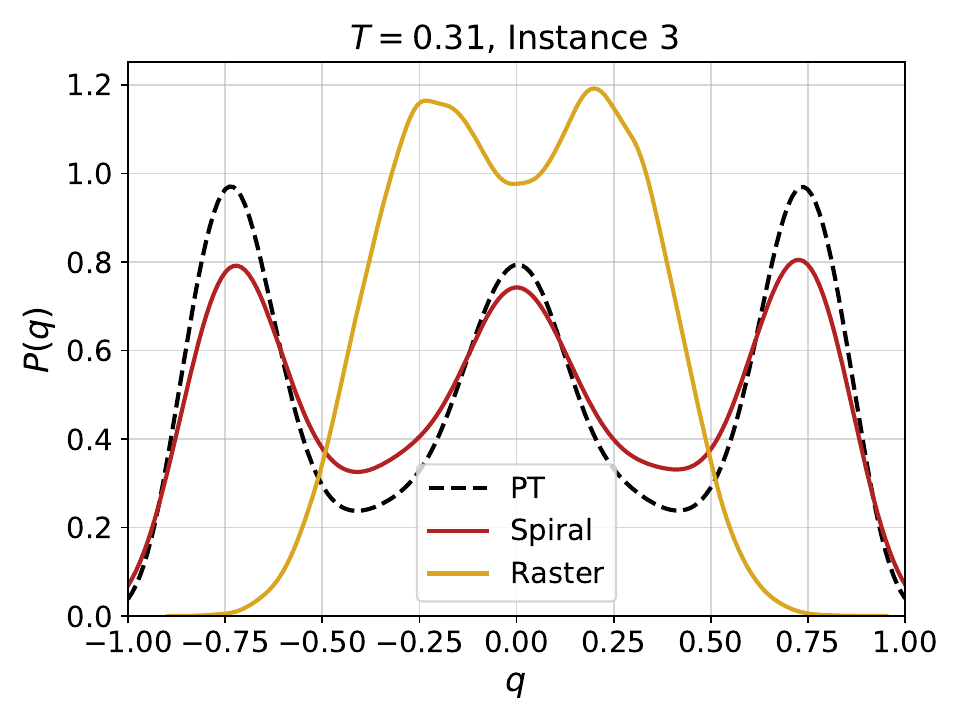}
        
    \end{minipage}
%
    \caption{Comparison of the probability distribution of the overlap obtained using different update sequences of the spins for $\ell = 10$. The black dashed line is obtained using data gathered via Parallel Tempering (PT), while solid color lines are obtained using $\Nf$ with different sequence updates. \textit{Spiral:} spins are update in a following a spiral going from the outside to the inside of the lattice; \textit{Raster:} spins are updated row by row from left to right. It appears as the spiral update performs much better than the raster update for the $\Nf$ architecture in equal training conditions, something that is not observed when training, for instance, the MADE architecture. This effect could be related to the locality characteristics of the $\Nf$ architecture.}
    \label{fig:PqSI3}
\end{figure*}

\pagebreak

\section{Cross-entropy plots in lin-lin and lin-log scale}

\begin{figure*}[!th]
    \centering
    \begin{minipage}[b]{0.45\textwidth}
        \centering
        \includegraphics[width=\textwidth]{Grafici/KL_diffL.pdf}
        
    \end{minipage}
    \hfill
    \begin{minipage}[b]{0.45\textwidth}
        \centering
        \includegraphics[width=\textwidth]{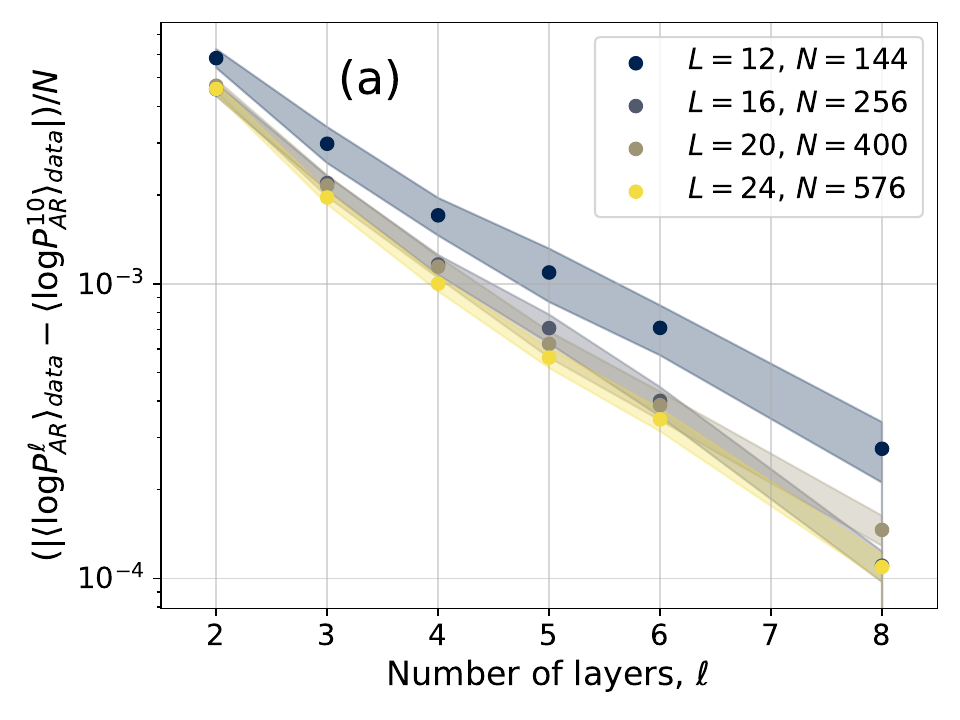}
        
    \end{minipage}
    \caption{Absolute difference between $\langle \log P^\ell_\text{NN} \rangle_\text{data}$ at various $\ell$ and at $\ell = 10$ for $T = 1.02$ and different system's sizes, in lin-lin and lin-log. Data are averaged over ten different instances of the disorder and the colored areas identify regions corresponding to plus or minus one standard deviation.}
    \label{fig:KL_diffL_diffscales}
\end{figure*}

\end{document}